\documentclass[manuscript]{aastex}
\usepackage{rotating}

\slugcomment{Submitted to Icarus}

\shorttitle{VLT/SINFONI observations of Uranus' 2014 bright storm}
\shortauthors{Irwin et al.}

\begin{document}


\title{Spectral analysis of Uranus' 2014 bright storm with VLT/SINFONI}


\author{P. G. J. Irwin, L.N. Fletcher, P.L. Read, D.Tice }
\affil{Department of Physics, University of Oxford, Parks Rd, Oxford OX1 3PU, UK.}
\email{irwin@atm.ox.ac.uk}

\author{I. de Pater}
\affil{University of California, Berkeley, CA 94720, USA.}

\author{G.S. Orton}
\affil{Jet Propulsion Laboratory, California Institute of Technology, 4800 Oak Grove Drive, Pasadena, CA 91109, USA.}

\author{N.A. Teanby}
\affil{School of Earth Sciences, University of Bristol, Wills Memorial Building, QueenÕs Road, Bristol, BS8 1RJ, UK}

\and

\author{G.R. Davis}
\affil{Square Kilometre Array Organisation, Jodrell Bank Observatory, Lower Withington Macclesfield, Cheshire, SK11 9DL, UK}


\begin{abstract}
An extremely bright storm system observed in Uranus' atmosphere by amateur observers in September 2014 triggered an international campaign to observe this feature with many telescopes across the world. Observations of the storm system in the near infrared were acquired in October and November 2014 with SINFONI on ESO's Very Large Telescope (VLT) in Chile. SINFONI is an Integral Field Unit spectrometer returning  $64 \times 64$ pixel images with 2048 wavelengths and uses adaptive optics. Image cubes in the H-band (1.43 -- 1.87 $\mu$m) were obtained at spatial resolutions of $\sim 0.1$\arcsec  per pixel. 

The observations show that the centre of the storm feature shifts markedly with increasing altitude, moving in the retrograde direction and slightly poleward with increasing altitude. We also see a faint `tail' of more reflective material to the immediate south of the storm, which again trails in the retrograde direction. The observed spectra were analysed with the radiative transfer and retrieval code, NEMESIS (Irwin et al., JSQRT 109, 1136, 2008). We find that the storm is well-modelled using either two main cloud layers of a 5-layer aerosol model based on \cite{sromovsky11} or by the simpler two-cloud-layer model of \cite{tice13}. The deep component appears to be due to a brightening (i.e. an increase in reflectivity) and increase in altitude of the main tropospheric cloud deck at 2 -- 3 bars for both models, while the upper component of the feature was modelled as being due to either a thickening of the tropospheric haze of the 2-layer model or a vertical extension of the upper tropospheric cloud of the 5-layer model, assumed to be composed of methane ice and based at the methane condensation level of our assumed vertical temperature and abundance profile at 1.23 bar. We also found this methane ice cloud to be responsible for the faint `tail' seen to the feature's south and the brighter polar `hood' seen in all observations polewards of $\sim 45^\circ$N for the 5-layer model.

During the twelve days between our sets of observations the higher-altitude component of the feature was observed to have brightened significantly and extended to even higher altitudes, while the deeper component faded.

\end{abstract}





\section{Introduction}

Although NASA's Voyager 2 spacecraft found Uranus to have a relatively featureless atmosphere during its fly-by of the planet in 1986, improved ground-based observations with ever-larger telescopes employing adaptive optics techniques have  revealed the atmosphere of Uranus to be much more dynamically active than that seen by Voyager 2. In addition to larger telescopes and better imaging, a new class of instruments, Integral Field Unit (IFU) spectrometers, have become available, such as the SINFONI instrument at the European Southern Observatory's (ESO) Very Large Telescope (VLT), the NIFS instrument at Gemini-North, and the OSIRIS instrument at Keck II. Such instruments can simultaneously map planets like Uranus at thousands of wavelengths with spectral resolving powers in excess of $R=\lambda/\Delta\lambda = 1000$. 

Since the mid 1990s several discrete clouds were seen in Uranus' atmosphere, generally at mid-latitudes, which became more frequent in the years leading up to the planet's northern spring equinox in 2007 \citep[e.g.][]{sromovsky07, sromovsky09}. Since equinox, Uranus continued to remain dynamically active, although overall cloud activity decreased. A notable exception to this was the detection of a bright spot near 25$^\circ$N in November 2011 \citep{sromovsky12}. 

Uranus cloud observations obtained with the Keck Telescope in August 2014  \citep{dePater15} revealed an amazingly active Uranus; numerous cloud features were observed, amongst them the brightest cloud ever seen at a wavelength of 2.2 $\mu$m; this cloud (`Br' in their nomenclature) was seen at a latitude of $\sim15.5^\circ$N. These observations triggered a campaign by amateur astronomers and shortly thereafter a very bright cloud was observed in September 2014. For this feature to have been detected at visible wavelengths with amateur telescopes strongly suggests that it must have had substantial optical depth, suggesting a very significant convective event. 
Both from tracking the position of this feature and its latitude, it was discovered to have evolved not from the brightest `Br' feature observed by \cite{dePater15}, but from a smaller feature at a higher latitude ($\sim 33^\circ$N), `Feature 2' in de Pater et al.'s nomenclature. This feature was identified as the deepest atmospheric feature seen with the Keck telescope in August 2014, at a pressure of near 2 bar. It also had an intriguing `tail' trailing in the retrograde direction. The fact that the cloud had been observed by amateurs sparked huge international interest amongst the professional planetary astronomy community and a number of Directors' Discretionary Time (DDT) proposals were submitted to telescopes around the world and the highly unusual event triggered a Hubble Space Telescope (HST) Target of Opportunity program. However, narrow-band imaging (e.g. HST) is insufficient to probe fully the three-dimensional structure of this spectacular eruption since observations, although highly detailed, can be obtained at only a few wavelengths subject to different atmospheric absorption and thus sounding different discrete pressure levels. Here we report the results of VLT/SINFONI DDT observations of Uranus made on October 31$^{st}$ and November 11$^{th}$ 2014 which enabled us to map the feature at both high spatial \textbf {and}  spectral resolution and thus much more precisely constrain the feature's vertical and horizontal cloud structure, also partially revealing its temporal evolution.

\section{Observations}

To observe the cloud seen by the amateur astronomers (i.e., `Feature 2' seen by \cite{dePater15}) observations of Uranus were made with the SINFONI instrument at the European Southern Observatory (ESO) Very Large Telescope (VLT) in La Paranal, Chile.  SINFONI is an Integral Field spectrograph that can make use of adaptive optics to yield a spatial resolution of typically 0.1\arcsec. Each one of SINFONI's 32 slitlets is imaged onto 64 pixels of the detector, giving $64 \times 32$ individual spectra, each with 2048 wavelengths, which are usually doubled in the cross-slit-direction to give $64 \times 64$ pixel `cubes'. SINFONI has three pixel scale settings: 0.25\arcsec , 0.1\arcsec  and 0.025\arcsec  giving Fields of View (FOV) of  8\arcsec  $\times$ 8\arcsec, 3\arcsec  $\times$ 3\arcsec  and 0.8\arcsec  $\times$ 0.8\arcsec. Uranus was observed on October 31$^{st}$  and November 11$^{th}$ 2014 using the 0.1\arcsec pixel scale and the H--grism, which has a spectral resolution of $R=\lambda/\Delta\lambda \sim 3000$ and covers the wavelength range 1.436 -- 1.863 $\mu$m. Since the FOV at the 0.1\arcsec pixel scale was smaller than the apparent disc size of Uranus at this time ($\sim$ 3.7\arcsec) dithered, overlapping observations were recorded on a 2 $\times$ 2 grid, with additional observations centred on Uranus' disc. Two sets of observations (Table \ref{tbl-1}) were made per night, spaced by a couple of hours, to allow the cloud to be observed at more than one emission angle.

The data were reduced with the ESO VLT SINFONI pipeline, but correction for the stellar absorption features of the telluric standard star was made using the Spextool \citep{cushing04} package  \textit{xtellcor-general}, which uses the method of \cite{vacca03}. Photometric correction was achieved by integrating the  observations of the A0 standard stars (HD212874 or HD210780) across the entire FOV, observed either immediately before or after each set of Uranus observations, using the quoted 2MASS \citep{cutri03} H-magnitudes  of 8.917 and 8.255, respectively, and the 2MASS H-filter profiles. Geometric registration was done manually and planetocentric latitudes were assumed throughout. We also corrected for the airmass difference between each frame making up the planet mosaic and the single calibrating standard star observation.

\begin{figure}
\epsscale{.80}
\plotone{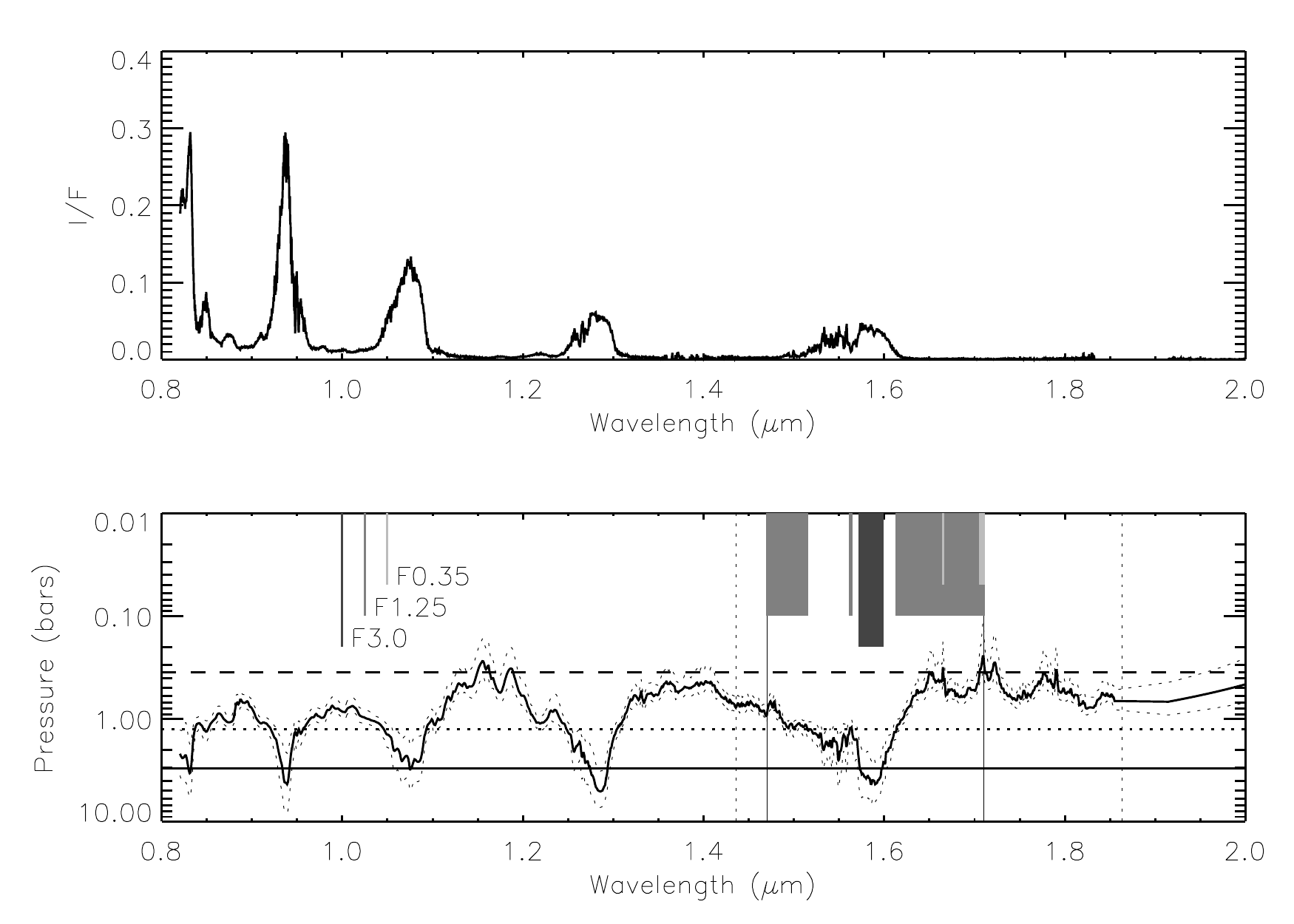}
\caption{Top panel shows a typical I/F spectrum of Uranus as observed by IRTF/SpeX. Bottom panel shows the pressure level in Uranus' atmosphere at which the two-way transmission to space for a cloud-free atmosphere (assuming the vertical profiles described in the text) is 0.5. Overplotted in the bottom panel are the pressure levels (dotted lines) for which the two-way transmission to space is 0.25 and 0.75, giving an indication of the vertical resolution of the observations at a single wavelength. Also overplotted in the bottom panel are the chosen cut-off pressures of 3, 1.25 and 0.35 bar. Continuum images (`F3.0') are averaged over all wavelengths where the two-way transmission to 3 bars exceeds 0.5. Medium-absorption and high-absorption images are averaged over all wavelengths where the two-way transmission at 1.25 and 0.35 bars is respectively \textbf{less} than 0.5, labelled respectively as `F1.25' and `F0.35'. The wavelengths selected by these filters in the wavelength range modelled (1.47 -- 1.71 $\mu$m, shown by the vertical solid lines) are indicated by the grey regions in the bottom panel of differing length and darkness; a key to these filter regions is indicated by the vertical bars in the top left of the bottom panel. Finally, the bottom panel also shows the total wavelength range of the VLT H-band SINFONI observations (1.436 -- 1.863 $\mu$m, vertical dotted lines).  \label{figpenetrate}}
\end{figure}

\begin{figure}
\epsscale{.80}
\plotone{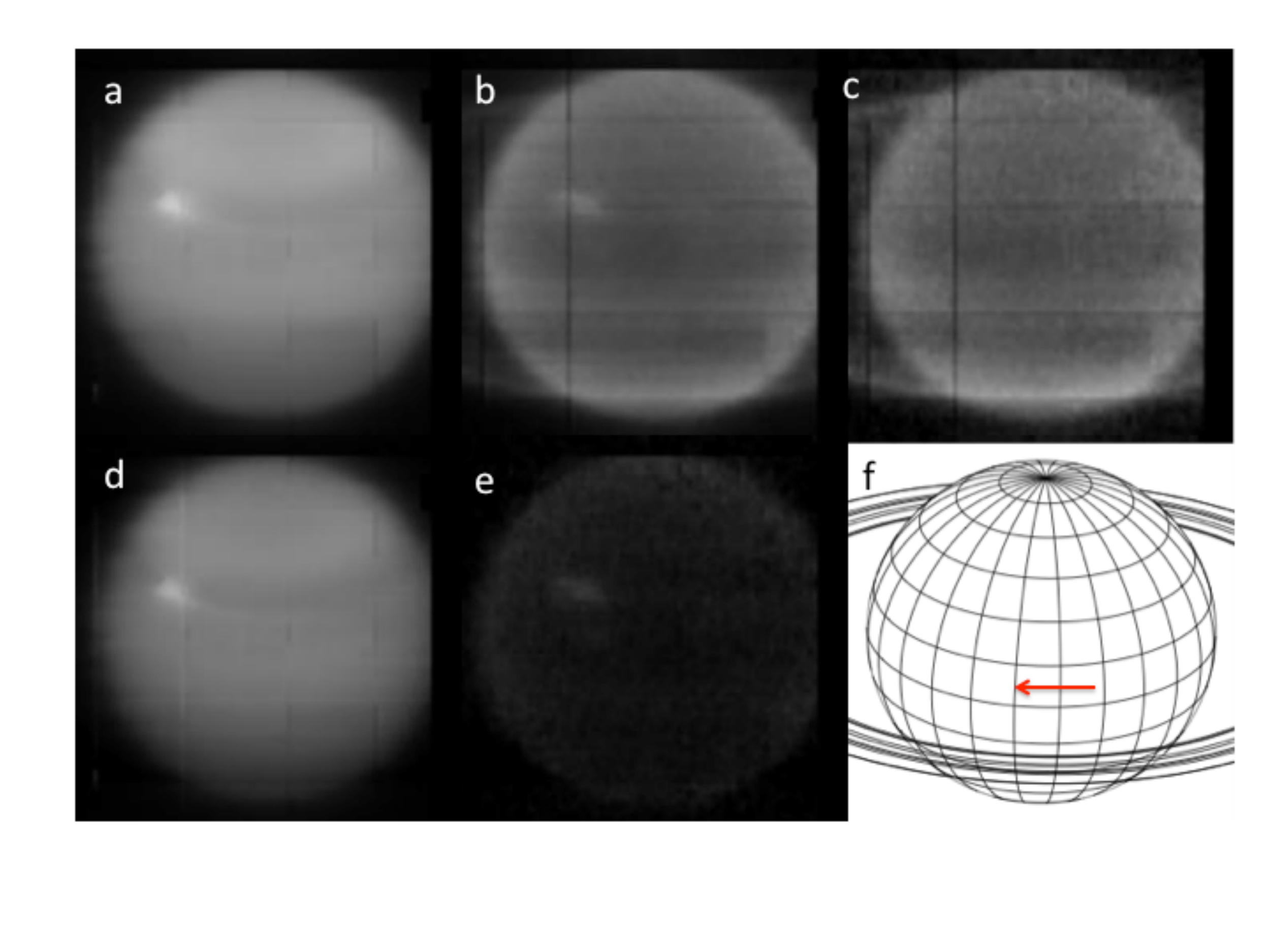}
\caption{Average of $1^{st}$ set of observations made on 31$^{st}$ October 2014 between 01:42 and 02:17 (UT). These images are the averaged mosiacs of all observed frames, where overlap between the frames creates the small striations seen. Uranus' north pole is at the top. The top row shows the appearance of Uranus in the different wavelength `filters'. Panel (a) shows the planet at wavelengths where the two-way transmission to space, for a cloud-free atmosphere, exceeds 0.5 at the 3-bar level (i.e. the `F3.0' filter). Panel (b) shows the planet at wavelengths where the two-way transmission to the 1.25 bar level is less than 0.5 (`F1.25'), while panel (c) shows the planet at wavelengths where the two-way transmission to the 0.35 bar level is less than 0.5 (`F0.35'). Uranus' epsilon-ring is visible in panels (b) and (c). The bottom row shows differences between the images to highlight the clouds at different levels. Panel (d) shows panel (a) minus panel (b) and shows the distribution of cloud reflectivity of the cloud decks at pressures $>  \sim1.25$ bar, while panel (e) shows panel (b) minus panel (c), showing the distribution of clouds roughly between 1.25 and 0.35 bar. Panel (f) shows a projection of Uranus' disc and rings for reference. Uranus' sense of rotation is indicated by the arrow in panel (f).\label{figoct31bw}}
\end{figure}

Figure \ref{figpenetrate} shows a typical reflectance spectrum of Uranus as measured by IRTF/SpeX\footnote{http://irtfweb.ifa.hawaii.edu/~spex/IRTF\_Spectral\_Library/}, together with the pressure level at which the two-way transmission to space is 0.5 for a cloud-free atmosphere, assuming the standard atmospheric profile described in the next section. The main absorption features seen in Uranus' near-IR spectrum are formed by gaseous methane. At wavelengths of strong methane absorption sunlight cannot penetrate very far and thus any light we see must have been reflected from hazes in the stratosphere. Conversely, in regions of weak absorption sunlight can penetrate to be reflected from clouds at the deepest levels. Hence, as Fig.\ref{figpenetrate} shows, such spectra allow us to determine the pressure level of clouds over a wide pressure range (0.1 -- 8 bar); it should be noted that while clouds lying at pressures greater than 8 bars cannot be detected as the atmosphere is too opaque, the presence of clouds lying at pressures less than 0.1 bar can be detected, but not their pressure level since the transmission to space is effectively unity at all wavelengths in this spectral region for pressures $< 0.1$bar. Hence, for such clouds/hazes we can measure their spectral reflectivity, but can only infer that they must lie at a pressure $<  0.1$ bar. In this paper, we present many false-colour plots, which show the distribution of deep, intermediate and high clouds/hazes. To generate these false-colour plots, we formed selected spectral averages to emulate filters targetting specific altitude levels. To map the deepest clouds we only use wavelengths where the two-way transmission to space (in a cloud-free atmosphere) exceeds 0.5 at the 3-bar level. We shall call this the `F3.0' filter. To map the intermediate clouds we choose only those wavelengths where the two-way transmission to space is \textbf{less} than 0.5 at the 1.25-bar level (`F1.25' filter). To map the highest clouds we choose only those wavelengths where the two-way transmission to space is less than 0.5 at the 0.35-bar level (`F0.35' filter). The wavelengths covered by these three `filters' in the 1.47 -- 1.71 $\mu$m range, which is the range covered by our radiative transfer modelling discussed in the next section, are shown in Fig.\ref{figpenetrate}. The appearance of Uranus recorded from 01:42 -- 02:17 on October 31$^{st}$ 2014, using the 0.1\arcsec pixel scale in these three `filters'  is shown in the top row of Fig.\ref{figoct31bw}. It should be noted that the `F0.35' image is formed from an average of the data at only a few wavelengths and thus has considerably lower signal to noise than the other two images. The bottom row shows the differences between these images, highlighting the cloud density at low and medium altitudes (the F0.35 map already shows the cloud density at high altitudes), and also shows the aspect of Uranus at the time of observation. Figure \ref{figsumfalsecol} presents a false colour representation of these images, together with those observed at the other times listed in Table \ref{tbl-1} and shows the distribution of low, medium and high altitude clouds/hazes. Figure \ref{figregionselect} shows the two cubes chosen for further analysis in this paper and shows the regions selected for radiative transfer modelling.

Although the observations on 11$^{th}$ November are slightly less well spatially resolved than those observed on 31$^{st}$ October, a number of key points are immediately clear from both sets of observations: 1) the centre of the cloud at the deepest levels is consistently to the left (i.e. in the prograde direction) and slightly below (i.e. to the south of) the intermediate clouds; 2) the intermediate level clouds are brighter on 11$^{th}$ November 2014, suggesting that clouds at this level have thickened and/or become more scattering in the two weeks since 31$^{st}$ October; 3) to the south and right (retrograde direction) of the storm cloud appears a faint `tail' of brighter material, also seen by \cite{dePater15}, running roughly along a line of constant latitude (this is more easily seen in the continuum image of Fig.\ref{figoct31bw}); and 4) the storm (which resides at a latitude of 34.5$^\circ$N) sits in a darker than average belt, lying between a medium-bright equatorial region and a significantly brighter `hood' polewards of 40 -- 50$^\circ$N. Observations on both nights also show very faint clouds `shadowing' the main features $\sim 60^{\circ}$ longitude away in the retrograde direction, but it is difficult to discern much from these features in these observations.

The images shown in Fig.\ref{figsumfalsecol} are mosaicked ($2\times2$ mosaics with a 5$^{th}$ frame centred on the disc) averages over a time span of $\sim 35$ minutes (Table \ref{tbl-1}). Assuming Uranus' bulk rotational period of 17h14m, Uranus rotates through $12.2^\circ$ during this time, which is approximately half the longitudinal extent of the cloud feature.  To capture a less rotationally distorted view, albeit at the expense of increased noise, Fig.\ref{figfalsecolsingle} shows the best individual frames targeting the storm in each average. Here we see the same key features and the reduced rotational smear is obvious. 

The apparent change in intermediate level cloud density suggested by these observations could potentially be due to differences in radiometric calibration between the two sets of observations. However, Fig.\ref{fignov11bw} shows the individual wavelength-selected images used to make up the false colour image of the second observation on 11$^{th}$ November, and here we can see that the intermediate-level clouds really are brighter relative to the background disc brightness. Furthermore, we can see traces of the cloud in the upper-level F0.35 image (i.e. where the two-way transmission to space at the 0.35-bar level for a cloudless atmosphere is less than 0.5), showing that the intermediate-level cloud is not only thicker, but extends to higher altitudes as well. Finally, the observations on 11$^{th}$  November clearly have lower spatial resolution than the 31$^{st}$ October images, which should have the effect of lowering the brightness of pixels containing storm features, by blurring their contribution into neighbouring pixels, rather than raising the brightness as is apparent here. 

\section{Radiative Transfer and Retrieval Analysis}

To analyse these data, the spectra in each pixel were first smoothed to an intermediate resolution, similar to that of the IRTF-SpeX instrument with a triangular-shaped instrument function with Full Width Half Maximum (FWHM) = 0.002 $\mu$m, giving a spectral resolution of $R \sim 775$. Although this sacrificed spectral resolution, it greatly increased our computation speeds and improved the signal-to-noise (SNR) ratio. This choice was justified by our previous high spectral resolution analysis of Uranus spectra \citep{irwin12}. From this analysis we concluded that, for cloud parameter retrievals, the lower  IRTF-SpeX resolution was the best compromise between computational efficiency, vertical resolution and SNR. Smoothing the spectra further would lower the noise levels (i.e. increase the SNR), but degrade the vertical resolution, while a higher spectral resolution greatly increases computation time of the radiative transfer code (which uses a Matrix-Operator multiple scattering model), while not greatly increasing the vertical resolution due to the lower SNR.

The temperature and abundance profile assumed in this study was the same as that used by \cite{irwin15}. The temperature profile was based on the `F1' profile determined by \cite{sromovsky11} which has an He:H$_2$ ratio of 0.131 and assumes a 0.04\% mole fraction of neon and a deep CH$_4$ mole fraction of 4\% (reducing with height), which \cite{kark09} found to be most appropriate for latitudes equatorwards of 45$^\circ$N,S.
 
These data were analysed using methane line data from the WKMC-80K line database \citep{campargue12} in the same way as described by \cite{irwin12}.  The spectra were fitted with the NEMESIS \citep{irwin08} radiative transfer and retrieval code, using a correlated-k radiative transfer model \citep{lacisoinas91} and methane k-tables derived from the WKMC-80K line data, assuming the IRTF-SpeX triangular instrument function with FWHM =  0.002 $\mu$m. Given the spectral coverage of the \cite{campargue12} line data and to keep clear of the strong telluric absorption band at 1.4 $\mu$m, the spectra were analysed over the range 1.47 -- 1.71 $\mu$m. These k-tables were computed using the hydrogen-broadened methane line shape of \cite{hartmann02}  (suitable for atmospheres where H$_2$ is the main constituent) and have a line wing cut-off of 350 cm$^{-1}$, which we previously found to give good fits to our Uranus and Neptune Gemini/NIFS observations \citep{irwin12}.  For this k-table, a CH$_3$D/CH$_4$ ratio of $3.6 \times 10^{-4}$, determined for Uranus by \cite{debergh86}, was assumed.  Although \cite{irwin12} revised this value downwards to $2.9 \times 10^{-4}$, the effect on cloud retrievals at IRTF/SpeX resolution is not significant. For H$_2$ -- H$_2$ and H$_2$ -- He collision-induced absorption (CIA) we used the coefficients of \cite{borysow91,borysow92} and \cite{zhengborysow95}. An equilibrium ortho/para-H$_2$ ratio was assumed at all altitudes and latitudes. Although \cite{conrath98} and \cite{fouchet03} show that the ortho/para-H$_2$ ratio actually varies quite significantly with both altitude and latitude, the effect on the spectra in this wavelength band is insignificant. In addition to H$_2$ -- H$_2$ and H$_2$ -- He CIA,  H$_2$ -- CH$_4$ and CH$_4$ -- CH$_4$ collision-induced absorption was also included \citep{borysowfrommhold86, borysowfrommhold87} as was Rayleigh scattering by the air molecules themselves. The spectra were simulated using a Matrix Operator multiple scattering code, based on the method of \cite{plass73}, with five zenith angles (with Gauss-Lobatto calculated ordinates and weights) and $N$ Fourier components to cover the azimuth variation, where $N$ is set adaptively from the viewing zenith angle, $\theta$, as $N$ = int$(\theta/3)$. The reference solar spectrum of \cite{fiorenza05} was used to simulate the solar flux.

To analyse these spectra two different cloud models were used: A) a simple two-layer model, favoured by \cite{tice13} and similar to that found to match Keck-II/OSIRIS observations by \cite{deKleer15}; and B) a more complex five-layer model similar to that described by \cite{sromovsky11}, and which \cite{irwin15} found to give a slightly better fit to IRTF/SpeX observations of Uranus than the two-layer Model A. The two-layer Model A has a thin tropospheric cloud (TC), based at around 2 -- 3 bar, and an extended tropospheric haze (TH) layer, based at 1 bar, with an adjustable fractional scale height.  To perform this calculation the reference temperature, pressure and abundance profiles were split into 39 levels equally spaced in log pressure between 10 bar and 0.003 bar. The more complex Model B has three thin clouds: a lower tropospheric cloud (LTC) based at 5 bar,  a middle tropospheric cloud (MTC) based at around 2 -- 3 bar, and an upper tropospheric cloud (UTC), assumed to be methane. Higher in the atmosphere are two haze layers: a tropospheric haze (TH) extending between 0.9 and 0.1 bar, and a stratospheric haze (SH) extending between 0.1 and 0.01 bar. Assuming that the bright storm clouds might be composed of methane ice, we fixed the UTC to the methane condensation level for the temperature-abundance profile assumed of 1.23 bar. However, instead of setting this cloud to be vertically thin, as assumed by \cite{sromovsky11} and \cite{irwin15}, we allowed it to be vertically extended, described by a variable fractional scale height. This model is summarised in Table \ref{tbl-2}. In this model, the reference temperature-pressure profile was split into 34 layers between 12 and 0.003 bar. The three lower thin clouds were modelled with one layer each (although the UTC was allowed to extend into the upper layers), while there were 5 layers below the LTC, 4 layers above the SH, and 4 layers each between the discrete clouds, all equally spaced in log(pressure). The TH and SH were each split into 5 layers each. This 5-layer model was subsequently found to provide a good fit to the observations, but the UTC occasionally extended beyond the tropopause, which was considered unphysical and would also lead to excessive reflectance in regions of strong methane absorption.  \cite{deKleer15} analysed Keck-II/OSIRIS observations of a `moderately bright' discrete cloud observed in Uranus' bright circumpolar band at 45$^\circ$N in both H and K bands in 2011, and found no trace of the feature in the K-band, which suggests such disturbances are confined below the tropopause. Hence, we introduced an additional parameterisation to our Model B to force the methane cloud density to reduce to zero at the tropopause level (at 0.1bar) with a tuneable degree of steepness governed by a pressure-dependent multiplicative factor: $1.0 - \exp(-((\log(p)-\log(0.1))/\alpha)^2)$, where the parameter $\alpha$ was fitted by the retrieval model and $p$ is the pressure (in bar). Although it is usually assumed that, as on Earth, the tropopause acts like a sort of lid, \cite{dePater14} argue that for Neptune, material in vortices may ascend through the tropopause to higher levels since the temperature-pressure profile is so isothermal in this region. Since the temperature-pressure profile of Uranus is estimated to be even more isothermal in the tropopause region, it is possible that the tropospheric `lid' constraint could also be broken at certain locations in Uranus' atmosphere. However, for the storm cloud considered here, there is very little trace at stratospheric altitudes, which suggests the vigour of this convective outbreak is not as strong as those occurring in Neptune's atmosphere.

As in the analysis of \cite{irwin15}, we made use of a novel retrieval technique where, in addition to the cloud opacity and vertical position parameters, we also retrieved the imaginary refractive index spectra of the cloud particles in some of the layers. These imaginary refractive index spectra were used in a Kramers-Kronig analysis to compute self-consistent extinction cross-section, single-scattering albedo and phase function spectra using Mie scattering, with the adjustment that we approximated the phase functions with combined Henyey-Greenstein functions. This adjustment was to eliminate phase function features peculiar to purely spherical particles, such as the `glory' and `rainbow', which are unlikely to be present for Uranian condensates as they are all predicted to be solid phase and thus almost certainly not spherical.  \cite{irwin15} found that such an approach, when applied to the particles in the TC (or MTC) led to a dramatically improved fit at the longwave edge of the 1.55 $\mu$m reflectance peak. To analyse these VLT Uranus spectra we adopted the same approach for the particles in the Tropospheric Cloud of Model A and Middle Tropospheric Cloud (MTC) of Model B, and the Tropospheric Haze (TH) particles in both models, assuming \textit{a priori} mean radii of 1.0 and 0.1 $\mu$m (with variance 0.05) respectively, and  \textit{a priori} refractive indices $1.4+0.001i$ at all wavelengths. When performing the Kramers-Kronig analysis, the real part of the refractive index was set to 1.4 at a wavelength of 1.6 $\mu$m. For Model B, following \cite{irwin15}, the scattering properties of the LTC were set to those empirically derived for the lower cloud by \cite{tice13}, while we used the methane refractive indices of \cite{martonchik94} for the UTC. The spectral properties of the SH were those described by \cite{sromovsky11}. \cite{irwin15} found that the complex refractive index spectrum of the TH was only loosely constrained by our H-band observations and hence, in the interests of computation time, the refractive index spectrum of the TH was left as \textit{a priori} for both Models A and B.

To assess the suitability of these cloud models, we initially analysed the first set of observations recorded on 31$^{st}$ October. We chose this set over the second set from that night since the feature was observed at lower zenith angles (making our multiple-scattering retrieval calculations significantly faster) and there was likely to be less systematic error due to the finite spatial resolution and mixing with space/near-limb pixels. We picked three points: 1) a region well away from the storm (and trailing cloud feature) at the same latitude, 2) the centre of the deeper cloud feature, and 3) the centre of the upper cloud feature. These points are indicated in Fig.\ref{figregionselect}. Figure \ref{specref1oct31} shows the fit we achieve with both the two-layer (Model A) and five-layer (Model B) models at Point 1. In these retrievals we found that we could not in general fit the observations to the precision of the random errors of the observed radiances computed by the VLT/SINFONI pipeline (in general, the results of an optimal estimation retrieval model are only reliable if $\chi^2/n \sim 1$). The discrepancies are believed to be caused by remaining errors in the assumed absorption coefficients and/or deficiencies in the assumed model parameterisation, but can be simplistically accounted for by adding `forward-modelling' noise to the observed spectra to account for uncertainties in the forward (i.e. radiative transfer) model. To do this, we added the forward modelling error spectrum previously estimated from Uranus IRTF spectra in this range \citep{tice13}.  With these forward modelling errors we can see that the quality of fit is extremely good in both cases, with $\chi^2/n \sim 1$ well below 1.0 for both models A and B.  Figure \ref{specompoct31} compares the measured spectra for all three points (including random measurement errors) in both linear and log space. As can be seen Point 2 has significantly higher radiance at 1.57 $\mu$m, but Point 3 (centre of upper cloud feature) has the highest radiance in the methane absorption band from 1.6 to 1.7 $\mu$m, indicating that the cloud lies at higher altitudes. Figure \ref{retrievedcloud2} compares the vertical cloud profiles (opacity per bar, computed at 1.6 $\mu$m) derived for the three test points for the simple 2-layer Model A, while Figure \ref{retrievedcloud5} shows the vertical cloud profiles derived using the more complex 5-layer Model B.  Both figures also show the integrated optical depth from space. For the simple 2-layer Model A, Fig.\ref{retrievedcloud2} shows that going from Point 1 to Point 2, the main difference is a thickening of both the Tropospheric Cloud (TC) and Tropospheric Haze (TH). However, the required thickening of the TH is near the base, since the Fractional Scale Height (FSH) of the TH for Point 2 is clearly less. Increasing the TH opacity at all altitudes would lead to excessive reflectance at methane absorbing wavelengths and thus here the model has increased the opacity, but reduced the FSH of the Tropospheric Haze. For Point 3, we see a further increase in opacity of the TH and a slight reduction for the TC relative to Point 2, but no obvious further change to the FSH of the haze. In contrast, Fig.\ref{retrievedcloud2} shows that the retrieved base pressure of the TC is almost identical for all three points with a  value of $\sim 2$ bars. For the more complex 5-layer Model B, Fig.\ref{retrievedcloud5} shows that for Point 1 the methane UTC is retrieved to be vertically thin, as it is assumed to be in the  \cite{sromovsky11} model. However, for Point 2 the UTC can be seen to be considerably vertically extended, and for Point 3 it easily reaches the tropopause at 0.1 bar, where its abundance is reduced to zero by the additional $\alpha$ cut-off parameter. Contrastingly, Fig.\ref{retrievedcloud5} shows that the retrieved base pressure of the MTC is almost identical for all three points with values of 1.7 -- 1.9 bar. The estimated $\chi^{2}/n$ values indicated that the fit is good for all three spectra, but to make this absolutely clear Fig. \ref{specref3oct31} shows our fit for Point 3 with both Models A and B, which is our worst-fitting case, demonstrating that we fit the observations very well with both these models.

For both Models A and B, the retrieved imaginary refractive index spectrum (and derived real component) of the Tropospheric Cloud (TC for Model A, MTC for Model B) was very similar at all locations and a typical example is shown in  Fig. \ref{figrefindex}. The required imaginary refractive index spectrum is very similar to that derived from IRTF/SpeX spectra by \cite{irwin15} in this spectral range. As we can see an increasing imaginary refractive index is required from 1.5 -- 1.65 $\mu$m, which has the effect of reducing the reflectance of the cloud particles over this range and drastically improves the fit to the longwave edge of the main H-band reflectance peak. 

Having tested our models on three individual cases, we then ran our retrieval models over a wider region of the storm feature, covering the storm cloud itself and also the trailing cloud feature. The region covered by this wide-area retrieval is indicated in Fig.\ref{figregionselect}. Our fitted cloud parameters for Model A are shown in Fig.\ref{fig2cloudoct31}, while those derived using Model B are shown in Fig.\ref{fig5cloudoct31}. Note that since this is an average of several individual frames, artefacts are apparent at the joins between them. This is particularly clear at pixels with $x$-value 7 and 30 and thus linear features at these positions are simply artefacts. 

For the two-layer Model A, Fig. \ref{fig2cloudoct31} suggests that the deep cloud feature (seen at continuum wavelengths, panel (a)) is formed at the main cloud deck (i.e. TC, panel (b)), while the opacity of the haze (panel e) is responsible for the upper component of the storm cloud, seen at methane absorbing wavelengths (panel d). The trailing feature and polar `hood' visible at continuum wavelengths is apparent in both the retrieved opacity and base pressure of the TC, and also in the retrieved imaginary refractive index of the TC particles at wavelengths longer than 1.55 $\mu$m (panels (h) and (i)). The retrieved fractional scale height of the TH is relatively featureless, but is lower in the vicinity of the storm cloud, where the opacity is greater. 

For Model B, Fig.\ref{fig5cloudoct31} suggests that the deep cloud feature is again formed at the main cloud deck (i.e. the MTC, panel (b)), while the opacity of the UTC (i.e. methane cloud, panel (e)) is responsible for the `trailing feature', and also the polar `hood'. The higher component of the feature is seen to be caused by the increased fractional scale height of the UTC (panel (f)), which leads to this cloud having significant opacity just below the tropopause in a small region near $x=10$, and so needs to be limited by the cut-off parameter $\alpha$, which is seen to locally peak at this point (panel (h)).  In this model, the retrieved refractive indices of the particles MTC show much less spatial variation than for the particles in the TC for Model A, as can be seen in panel (i). 

\begin{figure}
\epsscale{.80}
\plotone{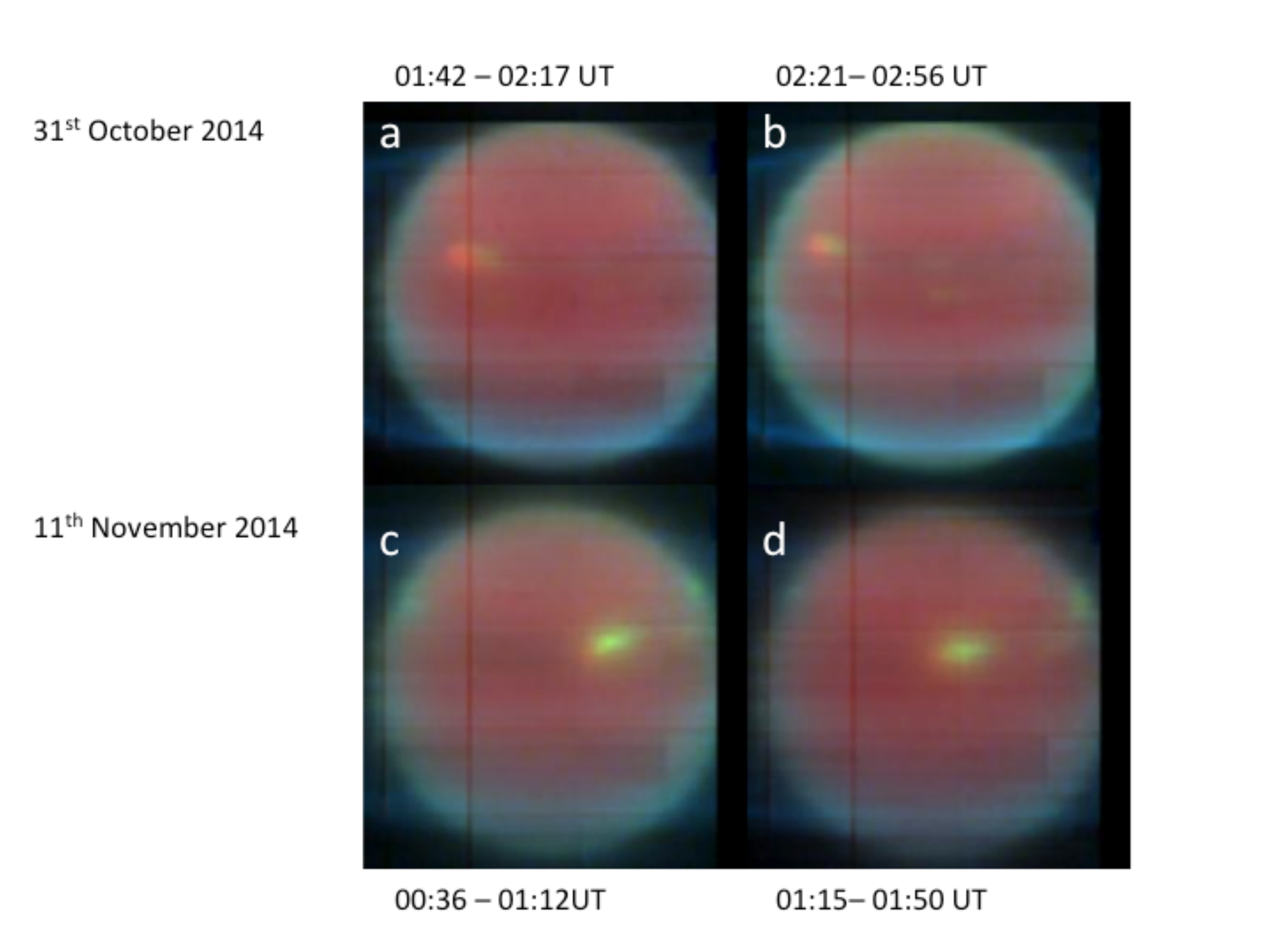}
\caption{False colour summary of observations. Observations of 31$^{st}$ October 2014 are on the top row, with panel (a) showing the average between 01:42 -- 02:17 and panel (b) showing the average from 02:21 -- 02:56 (UT), when the storm cloud is closer to the evening limb. In these false-colour images, red is the continuum F3.0 image (i.e. panel (a) in Fig.\ref{figoct31bw}), green is the F1.25 image where Trans$_{1.25bar} < 0.5$ (i.e. panel (b) in Fig.\ref{figoct31bw}) and blue is the F0.35 image where Trans$_{0.35bar} < 0.5$ (i.e. panel (c) in Fig.\ref{figoct31bw}). In this scheme, deep clouds appear red, intermediate clouds appear yellow and high hazes appear bluish.  Observations on 11$^{th}$ November 2014 are shown on the bottom row, with panel (c) showing the average between 00:36 -- 01:12, when the storm is nearer the morning limb and panel (d) showing the average from 01:15 -- 01:50 (UT). It can be seen that the centre of the deep cloud is offset from the centre of the cloud between 1.25 and 0.35 bar since the deep `red' cloud is consistently seen to the left of the overlying `yellow' cloud. 
\label{figsumfalsecol}}
\end{figure}

\begin{figure}
\epsscale{.80}
\plotone{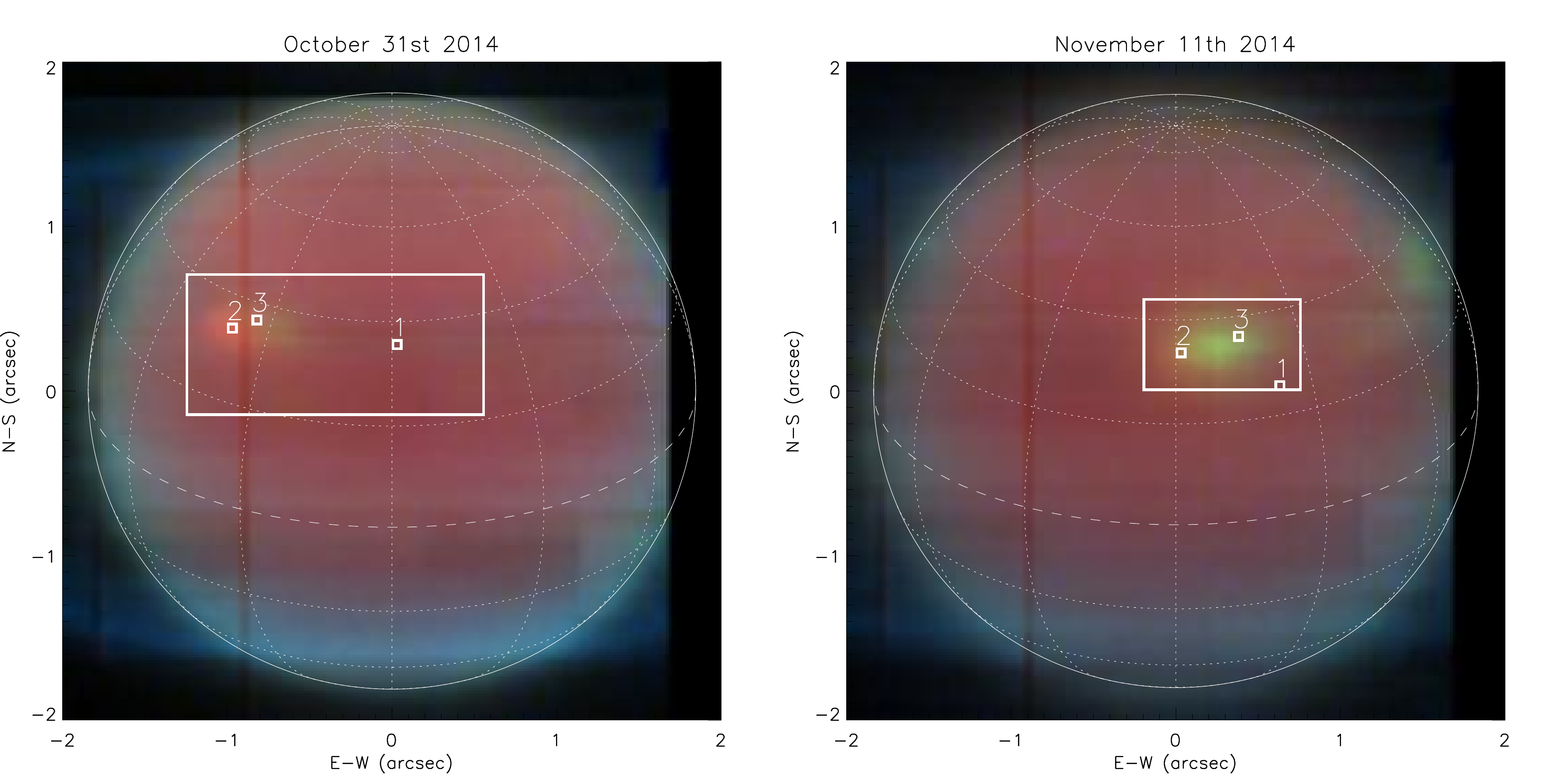}
\caption{The left and right hand panels reproduce panels a) and d) of Fig.\ref{figsumfalsecol} respectively, and indicate where spectra were extracted and analysed for the nights of 31$^{st}$ October 2014 (1$^{st}$ set)  and 11$^{th}$ November 2014 (2$^{nd}$ set) . Individual points are indicated, while the boxed areas indicate the regions mapped in our area retrievals, shown later in Figs.\ref{fig2cloudoct31} -- \ref{fig5cloudnov11}. Overplotted are latitude circles, spaced by $20^\circ$ and longitude circles spaced by $30^\circ$.\label{figregionselect}}
\end{figure}

\begin{figure}
\epsscale{.80}
\plotone{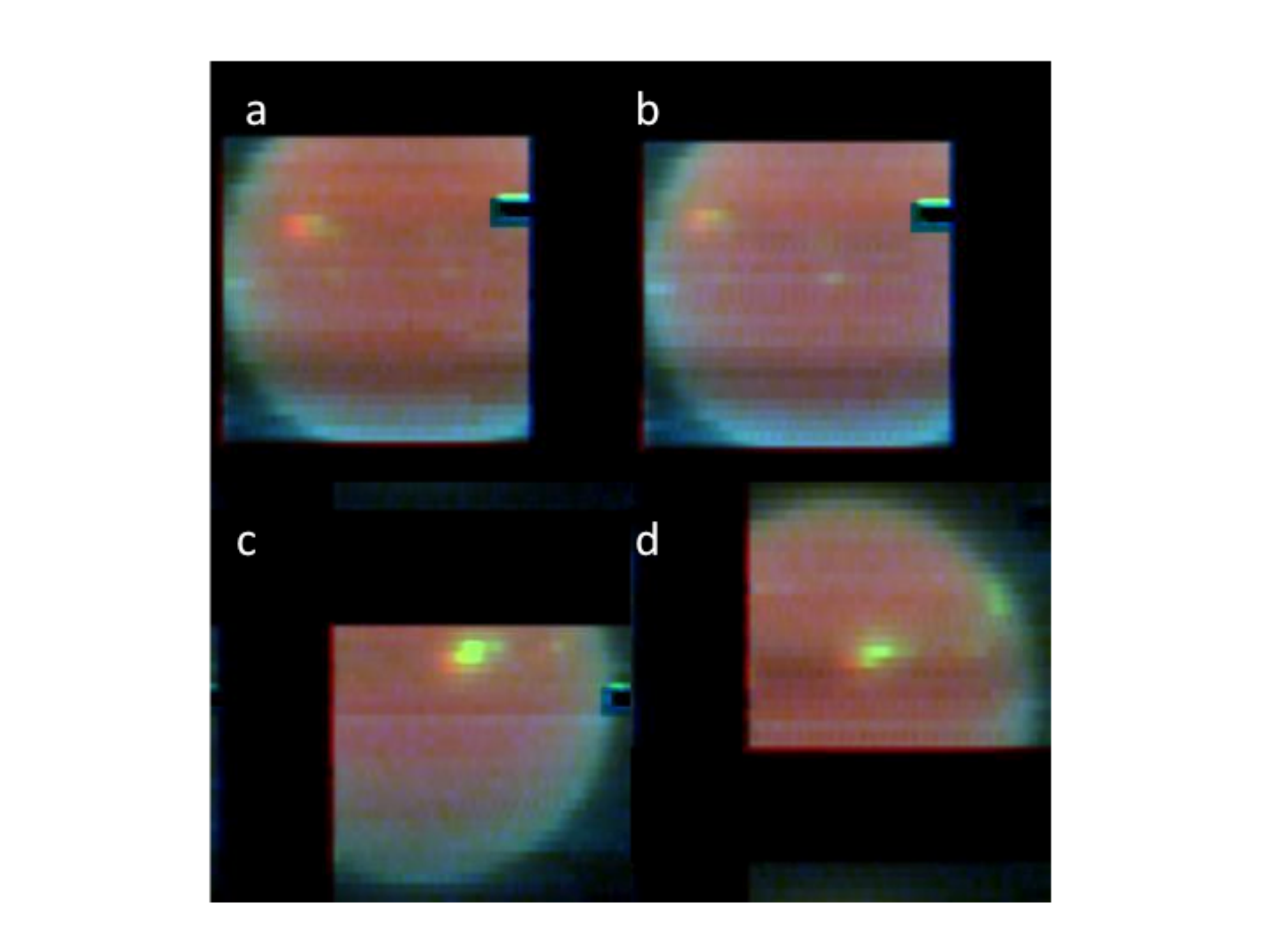}
\caption{As for Fig.\ref{figsumfalsecol}, except here we have used individual frames taken in the same observation blocks, to mitigate the smearing effect introduced by Uranus' rotation at the expense of reduced SNR. The increased brightness of the upper level clouds between 1.25 and 0.35 bars on 11$^{th}$ November 2014 is clear, as is the offset between the deep and upper components of the cloud feature.\label{figfalsecolsingle}}
\end{figure}

\begin{figure}
\epsscale{.80}
\plotone{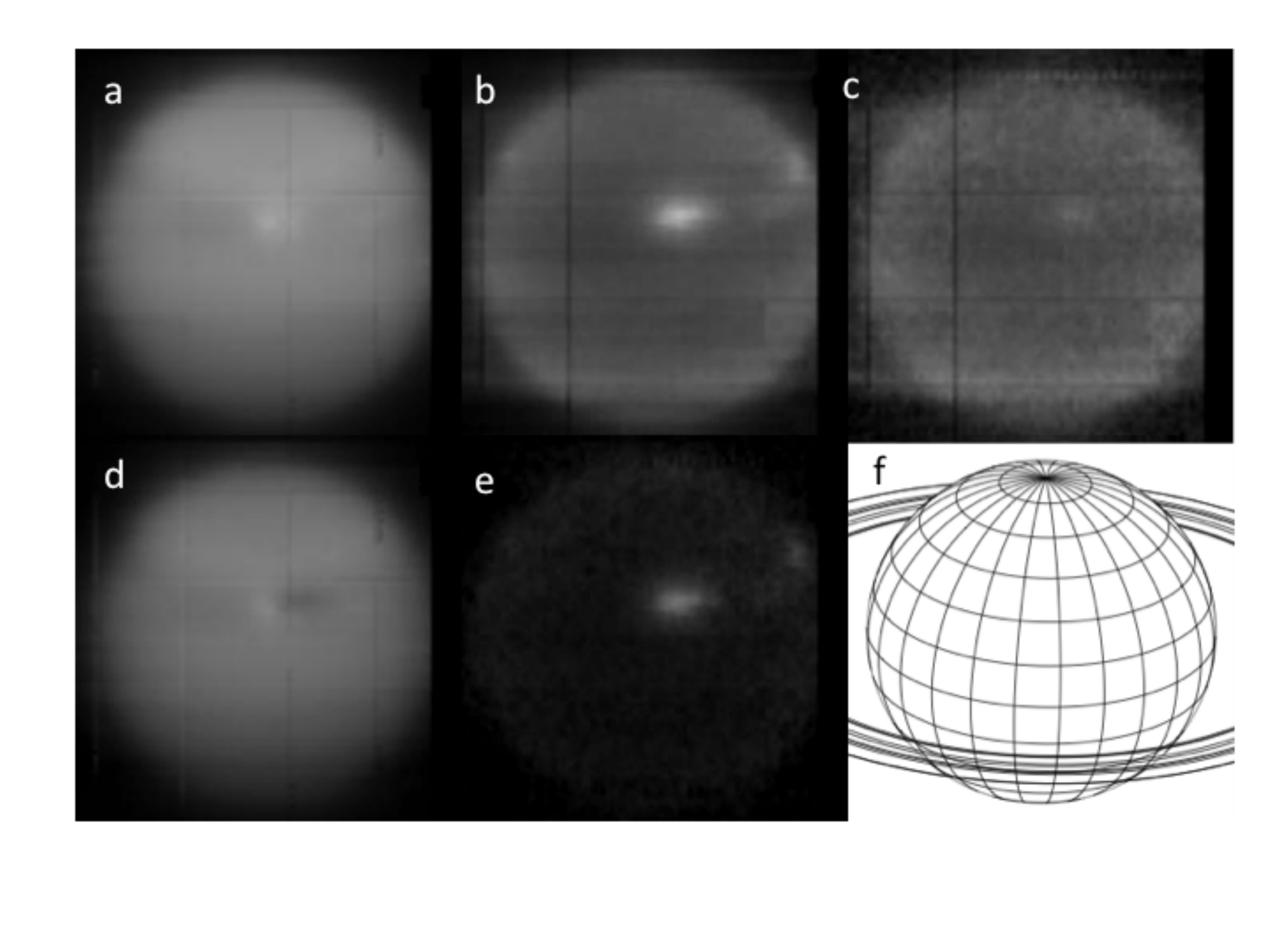}
\caption{As Fig.\ref{figoct31bw}, but showing the average of the 2$^{nd}$ set of observations made on 11$^{th}$ November 2014 between 01:15 and 01:50 (UT). The same brightness levels have been used in plotting these images and it can be seen that the cloud between the $\sim1$ and 0.35 bar levels is brighter here than on 31$^{st}$ October 2014. The cloud can also be seen faintly in panel (c), showing that the cloud really has reached a higher altitude than it had achieved on 31$^{st}$ October.\label{fignov11bw}}
\end{figure}

\begin{figure}
\epsscale{.60}
\plotone{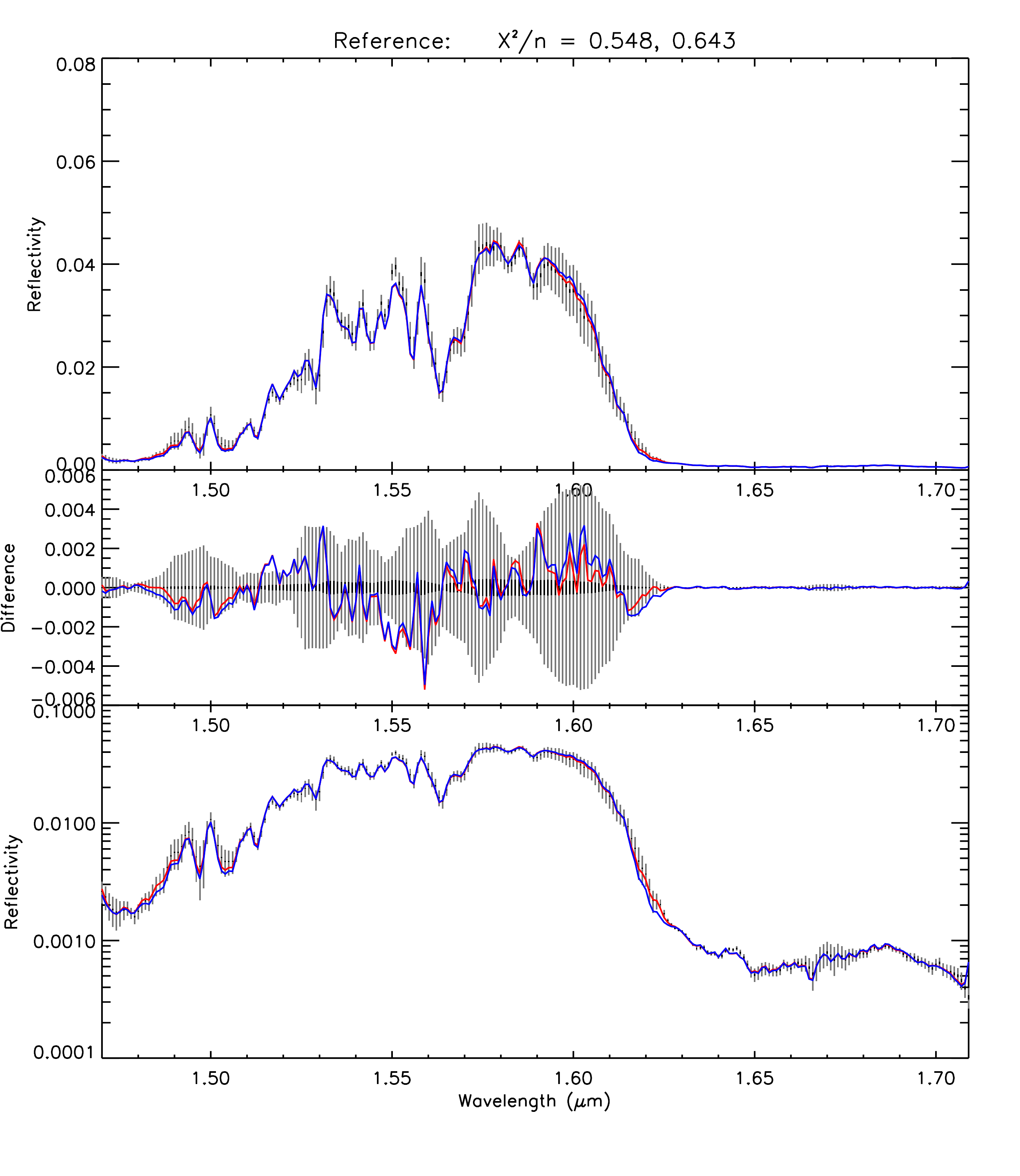}
\caption{Observed and fitted spectra at reference point 1 (Fig,\ref{figregionselect}), well away from the storm cloud on 31$^{st}$ October 2014. The blue line is the spectrum with the 2-cloud model (Model A), while the red line is that calculated with the 5-layer model (Model B). The $\chi^2/n$ values in the title indicate the closeness of fit for the 5-layer and 2-layer models respectively. Top panel compares the measured (grey error bars) and modelled spectra, the middle panel shows the difference between the measured and modelled spectra compared with the error bars, while the bottom panel compares the measured and modelled spectra on a log scale to highlight the low reflectivity regions. The black error bars that can be seen in the middle panel indicate the random noise level estimated by the VLT/SINFONI pipeline. The grey error bars indicated in all three panels include this random noise plus an additional estimate of forward modelling error employed by \cite{tice13} to account for errors in the line data and forward radiative transfer model. The values of $\chi^2/n$ quoted in the figure assume the latter, higher estimate of the uncertainties.\label{specref1oct31}}
\end{figure}

\begin{figure}
\epsscale{.80}
\plotone{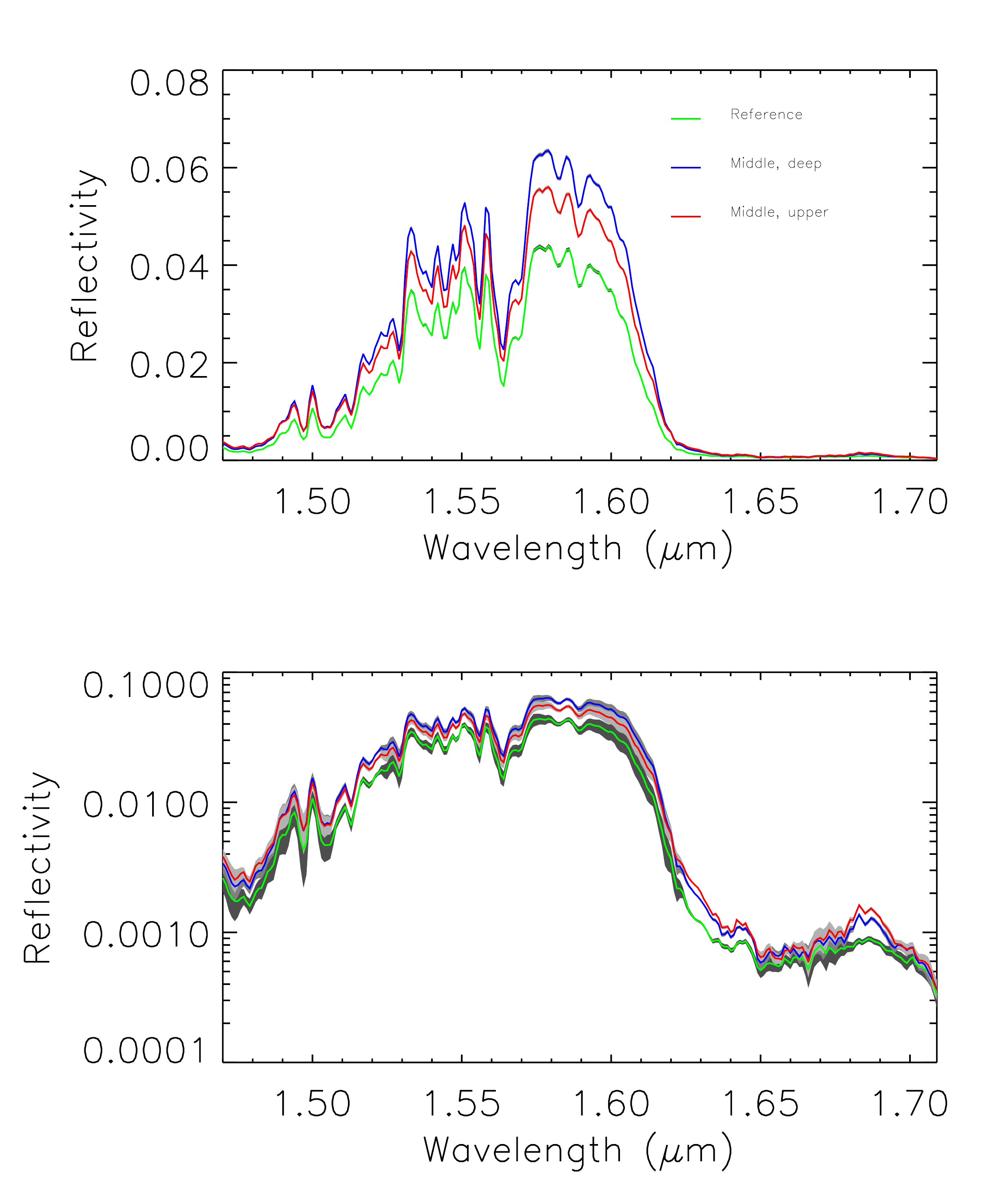}
\caption{Comparison of the measured spectra for the three sample points on 31$^{st}$ October 2014, indicated in Fig.\ref{figregionselect}: (1) reference point well away from storm; (2) middle of deeper cloud feature; (3) middle of upper cloud feature. The measurement errors estimated from the VLT/SINFONI pipeline are indicated by the grey regions. The top panel shows the reflectivity spectrum in linear units, while the bottom panel shows log(reflectivity) to accentuate the strongly absorbing spectral regions.   \label{specompoct31}}
\end{figure}

\begin{figure}
\epsscale{.80}
\plotone{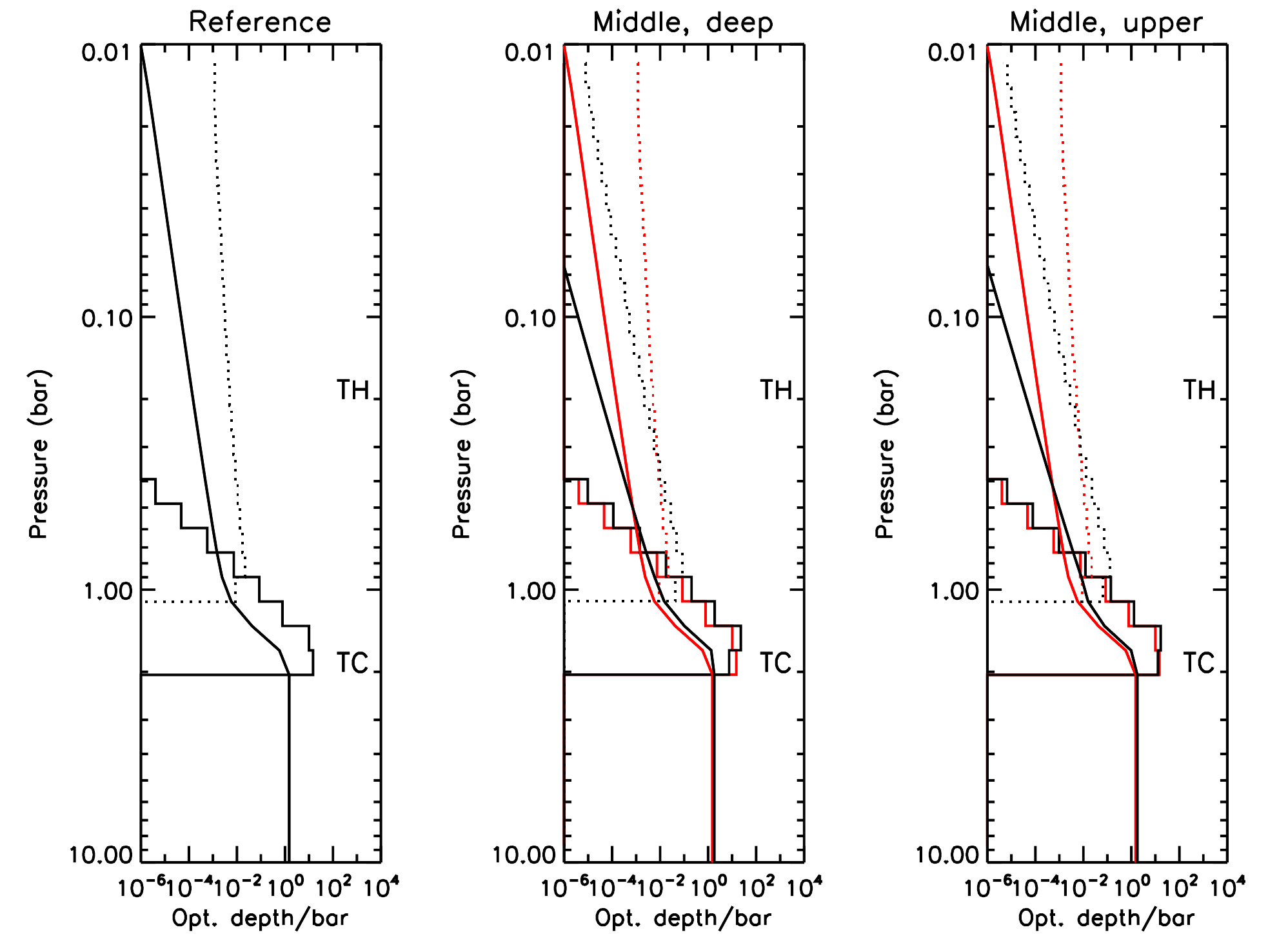}
\caption{Comparison of the three fitted cloud profiles retrieved from the three test points on 31$^{st}$ October 2014 for the simple 2-layer Model A: (1) reference point well away from storm; (2) middle of deeper cloud feature; (3) middle of the upper cloud feature. Optical depths are quoted at a wavelength of 1.6 $\mu$m. The continuous line increasing with pressure in all panels is the integrated optical depth. The retrieved profiles from the reference point (left hand panel) are reproduced in red in the middle and right hand panels to aid comparison. \label{retrievedcloud2}}
\end{figure}

\begin{figure}
\epsscale{.80}
\plotone{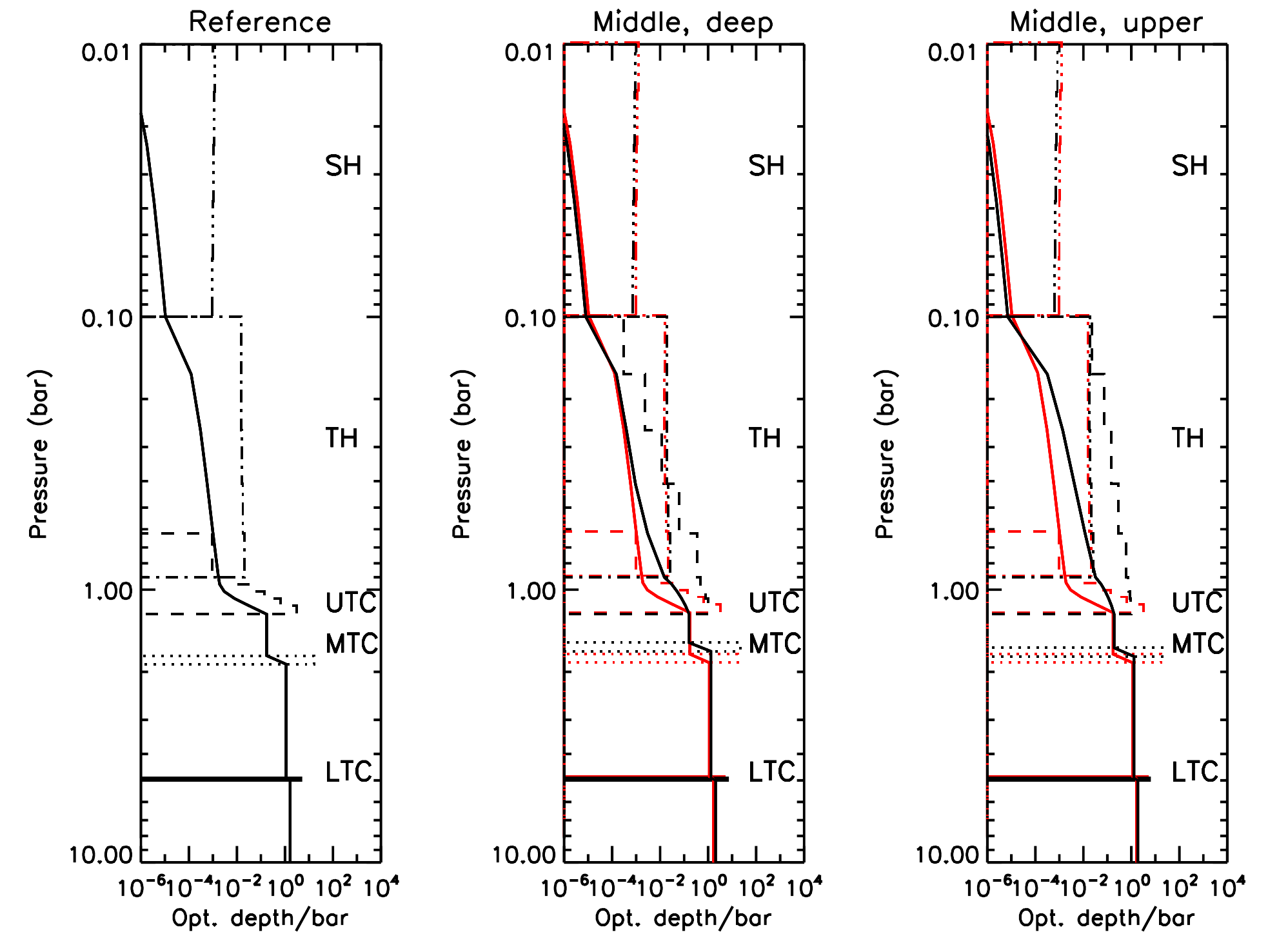}
\caption{As Fig.\ref{retrievedcloud2}, but showing a comparison of the three fitted cloud profiles retrieved from the three test points on 31$^{st}$ October 2014 for the more complex 5-layer Model B. Again, the optical depths are quoted at a wavelength of 1.6 $\mu$m and once again the retrieved profiles from the reference point (left hand panel) are reproduced in red in the middle and right hand panels to aid comparison.\label{retrievedcloud5}}
\end{figure}

\begin{figure}
\epsscale{.80}
\plotone{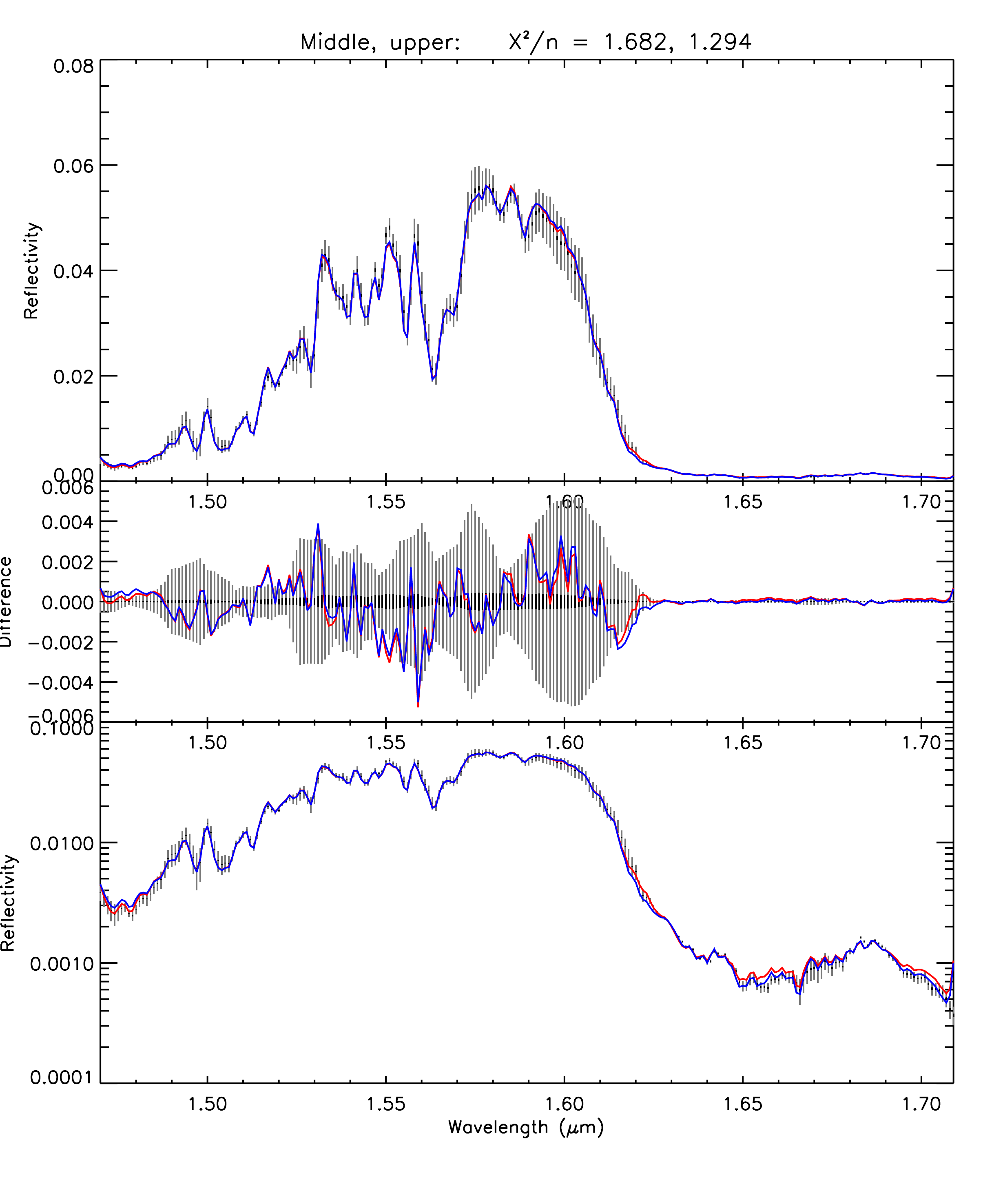}
\caption{As Fig.\ref{specref1oct31}, except for point 3 of the 31$^{st}$ October 2014 test retrievals (indicated in Fig.\ref{figregionselect}), in the middle of the higher cloud where our fit is poorest, showing that we still achieve an acceptable fit between the modelled and measured spectra. Again, the blue line is the spectrum computed with the 2-layer model, while the red line is that calculated with the 5-layer model.\label{specref3oct31}}
\end{figure}

\begin{figure}
\epsscale{.80}
\plotone{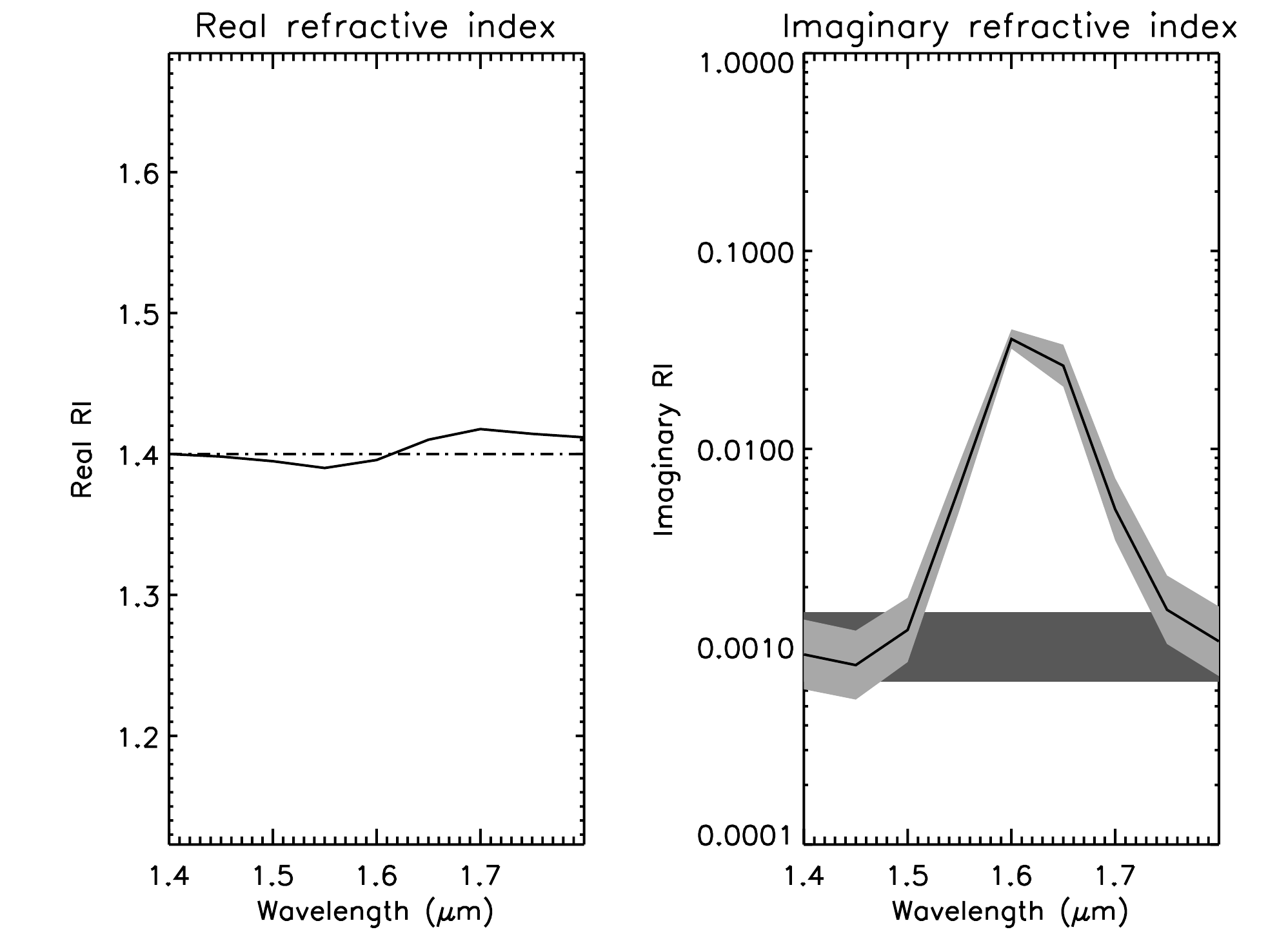}
\caption{Retrieved refractive index spectra of the Middle Tropospheric Cloud (MTC) at 2--3 bars for Point 1 of the 31$^{st}$ October 2014 test retrievals for the 5-layer Model B, showing adjustment necessary to properly fit the long-wave edge of the 1.55 $\mu$m peak. Very similar refractive index spectra were derived at all points and with both 2-layer and 5-layer models. In the left hand panel the \textit{a priori} real refractive index spectrum is the dot-dash line, while the fitted spectrum is the solid line. In the right hand panel, the \textit{a priori}  imaginary refractive index and errors are indicated by the dark grey region, while the fitted spectrum is shown by the solid line, with errors indicated in light grey.\label{figrefindex}}
\end{figure}

\begin{figure}
\epsscale{.80}
\plotone{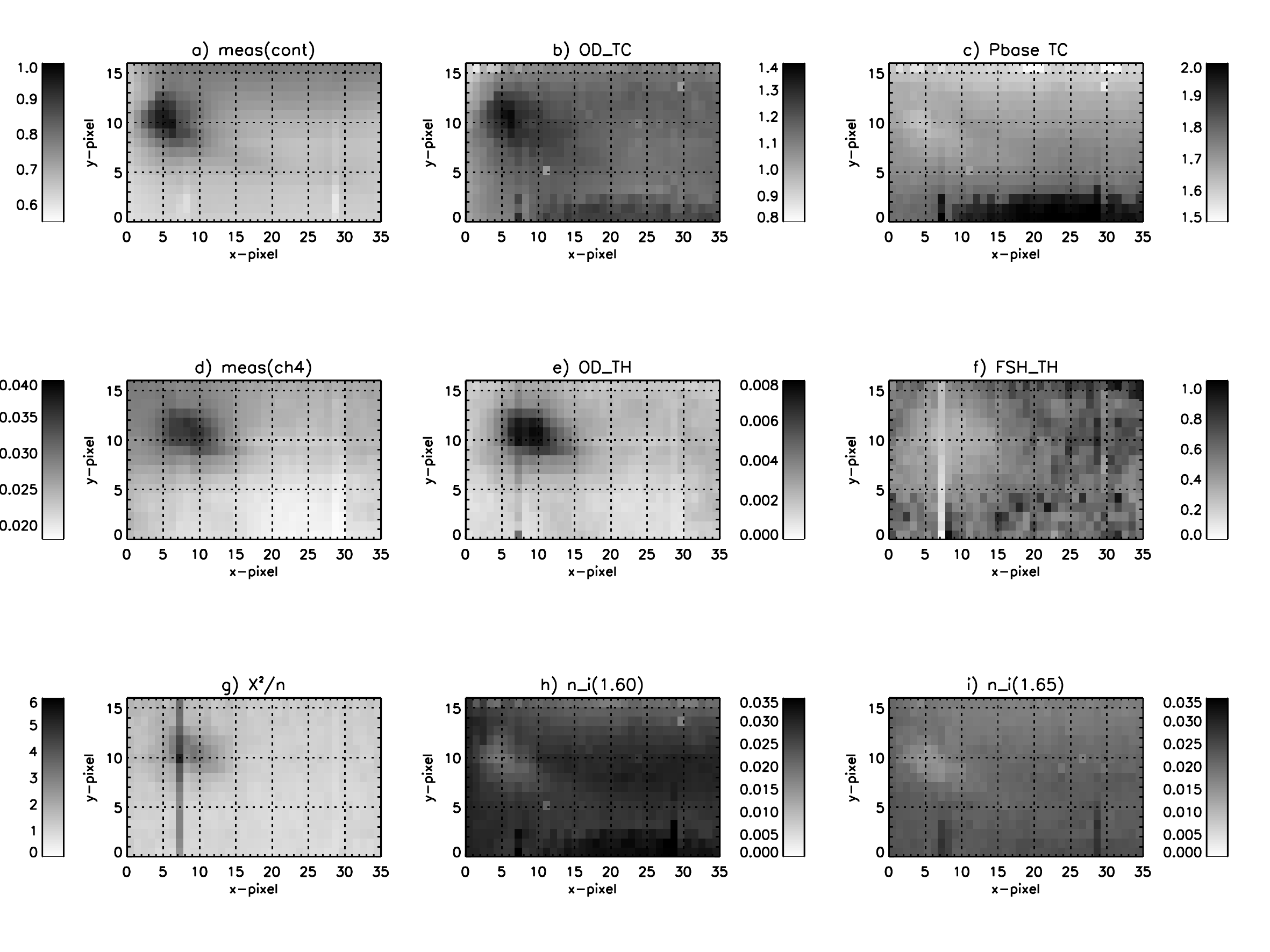}
\caption{Observed and fitted parameters for the first set of 31$^{st}$ October observations, fitted with the 2-layer Model A, over the area indicated in Fig.\ref{figregionselect}.  Panels a) and d) show the measured radiances in the continuum (average of 1.55 -- 1.62 $\mu$m) and medium methane absorption (average of 1.62 -- 1.65 $\mu$m) bands, while Panel g)  shows the variation of $\chi^2/n$. Radiance units are $\mu$W cm$^{-2}$ sr$^{-1}$ $\mu$m$^{-1}$. The other panels show: b) opacity of the Tropospheric Cloud (OD-TC); c) base pressure of TC (Pbase TC); e) opacity of the Tropospheric Haze (OD-Haze); f) Fractional scale height of Tropospheric Haze (FSH-Haze). Panels h) and i) show the imaginary refractive index of the MTC at two wavelengths (1.6 and 1.65 $\mu$m) where its value is well-constrained by the observations. \label{fig2cloudoct31}}
\end{figure}

\begin{figure}
\epsscale{.80}
\plotone{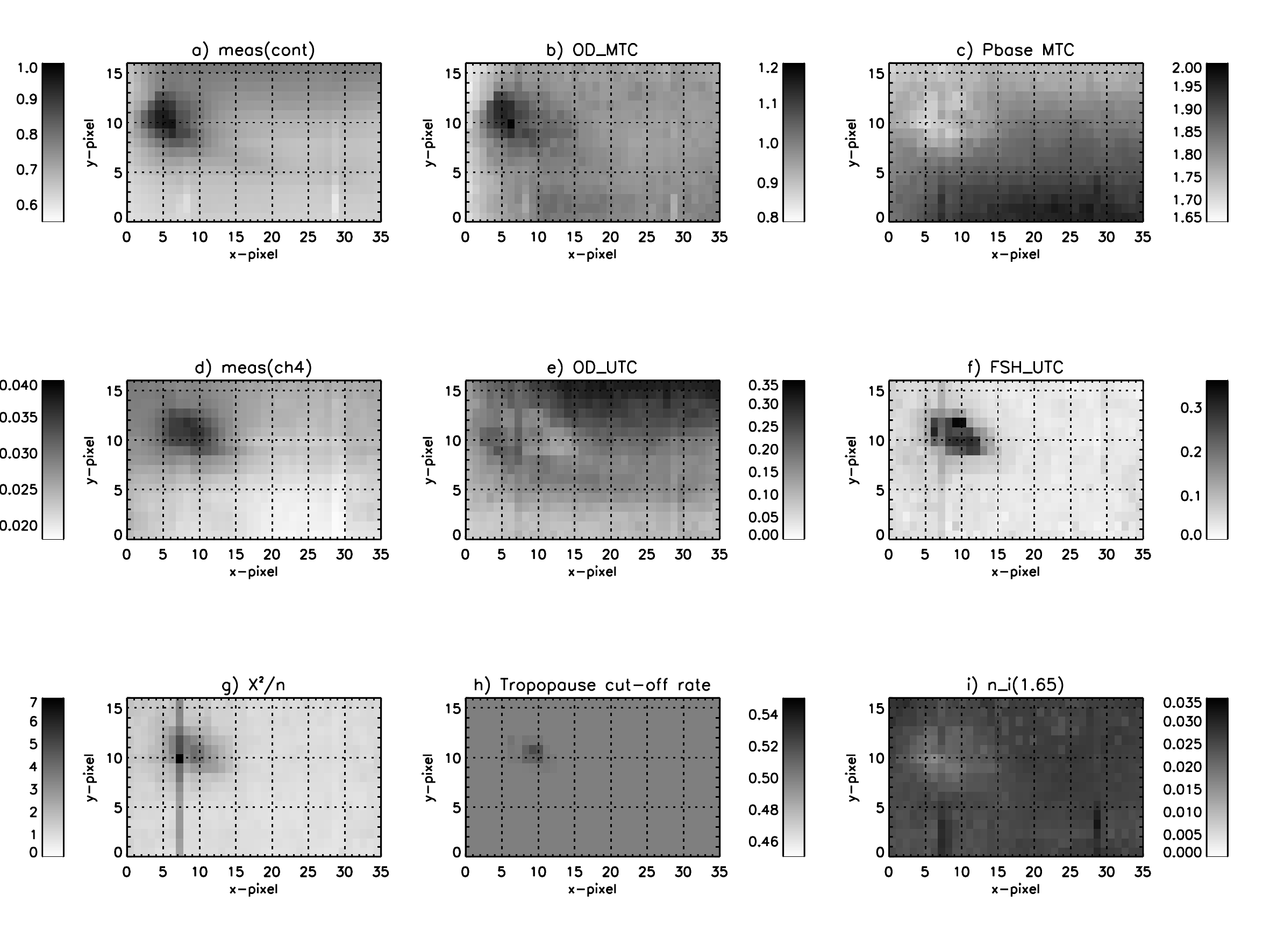}
\caption{Observed and fitted parameters for the first set of 31$^{st}$ October observations, fitted with the 5-layer Model B.  Panels a) and d) show the measured radiances in the continuum (average of 1.55 -- 1.62 $\mu$m) and medium methane absorption (average of 1.62 -- 1.65 $\mu$m) bands, while Panel g)  shows the variation of $\chi^2/n$. Radiance units are $\mu$W cm$^{-2}$ sr$^{-1}$ $\mu$m$^{-1}$. The area covered by these retrievals is indicated in Fig.\ref{figregionselect}. The other panels show: b) opacity of the Middle Tropospheric Cloud (OD-MTC); c) base pressure of MTC (Pbase MTC); e) opacity of the Upper Tropospheric Cloud (OD-UTC); f) Fractional scale height of UTC (FSH-UTC); h) Tropopause cut-off parameter, $\alpha$; g) imaginary refractive index of the MTC at 1.65 $\mu$m.\label{fig5cloudoct31}}
\end{figure}

The relatively narrow wavelength range of these H--band data meant that significant degeneracy was encountered between some elements of the 5-layer Model B since the spectral signature of changing the TH or SH opacity is rather similar and cannot easily be discriminated from each other or from the UTC fractional scale height when this parameter has a large value. The indistinguishability seen here in the H--band is in contrast to studies where a wider range of wavelengths has been considered \citep{irwin15, sromovsky11}, where the differing scattering spectra of these components allows them to be separated. Note that the \textit{a priori} effect of the SH at these wavelengths is particularly small as its opacity at 1.6 $\mu$m is only $\sim 10^{-5}$ and retrievals where its opacity were fixed were indistinguishable from those shown here. The expected opacity of the TH is somewhat larger at these wavelengths, however, ($\sim 10^{-3}$ at 1.6 $\mu$m) and there is a greater effect on the modelled spectrum. In particular, we found that when the fractional scale height of the UTC was large, its spectral signature became almost indistinguishable from the TH. This did not cause any problems here, where the UTC fractional scale height was mostly small ( $ < \sim 0.15$), but in our retrievals for  $11^{th}$ November, reported later, where larger values for this parameter were inferred for Model B, this led to instability, requiring us to fix the TH opacity at its mean value of 0.002 at 1.6 $\mu$m. To test whether fixing the TH opacity would affect the modelling of the $31^{st}$ October data we repeated these retrievals with the TH opacity fixed to 0.002. We found that the retrieved spatial distributions of the LTC, MTC and UTC parameters were the same as those shown here, which supports our initial assumption that, assuming Model B is correct, these transient opacity variations arise mostly from variations in the optical thickness of the MTC (i.e. the main cloud deck at  2--3 bar) and UTC (i.e. methane cloud), and also the vertical extent of the UTC. To determine reliable maps of the SH and TH opacity would require observations covering a wider wavelength range, which we do not have, and thus we fix them here to $\sim 10^{-5}$ and 0.002 respectively at 1.6 $\mu$m. However, while certain aspects of Model B cannot be constrained by these data alone, we prefer it over Model A since it has previously been shown to apply well over a large wavelength range (0.4 -- 2.0 $\mu$m) and its use of a separate methane cloud deck is in line with our expectations of Uranus' cloud structure. In particular, the upper part of the storm cloud seen in these observations clearly resides in the $\sim 1$ to 0.3 bar region, based upon the simple reflectivity differencing in Figs. \ref{figoct31bw} -- \ref{fignov11bw} and also from the retrievals with the 2-layer Model A, and its transient existence and discrete confinement suggests formation by condensation arising from convection rather than a `haze' feature, which might be thought to be longer-lived and less confined. It is also apparent, that while Model B can account for the storm cloud upper component, trailing feature and polar `hood' simply by increasing the opacity or vertical extent of the UTC, for the 2-layer Model A, considerable variation in position of all clouds and also the reflectivity of the TC particles is required. Hence, on balance, we believe the upper part of the storm cloud to be composed of condensed methane ice, but we cannot rule out alternative interpretations based on these observations alone since Models A and B fit the data equally well over the wavelength range observed. What is certainly clear, from the raw data, is that additional, transient opacity is needed at $\sim 1$ bar, which seems more consistent with a condensation cloud than a haze feature.

To test the effect of rotational `smearing', the best individual frame recorded in the first set of observations of October 31$^{st}$ (frame 17) was analysed with our retrieval model, focussing this time just on the storm region and the trailing cloud feature. The longitudinal spreading was reduced, as expected, but at the expense of a slightly poorer vertical resolution (caused by not adding several dithered frames leading to lower signal-to-noise ratios). We do not present these retrievals here, but from them the same general conclusions were drawn, assuming either Models A or B.

\begin{figure}
\epsscale{.80}
\plotone{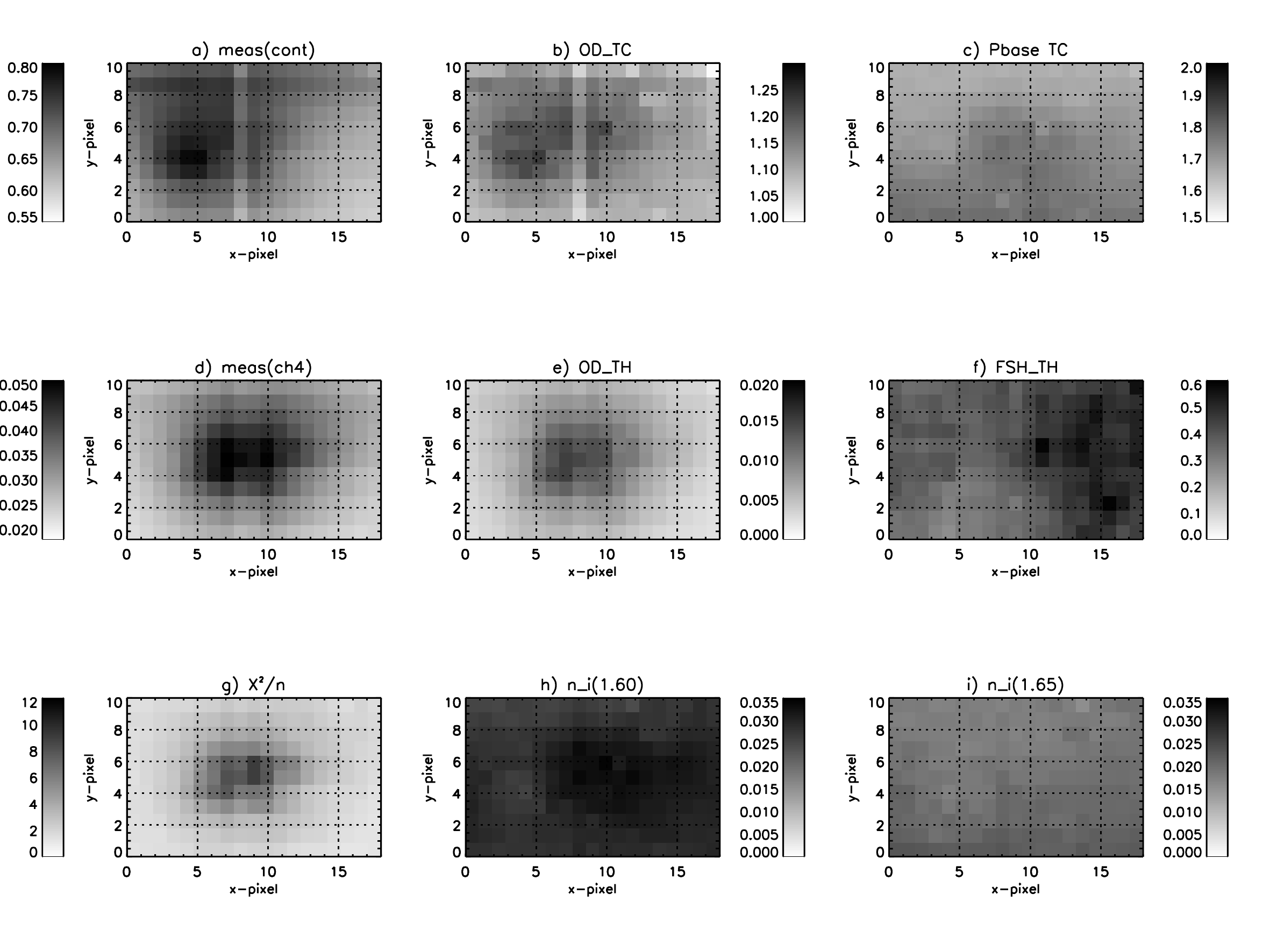}
\caption{As Fig.\ref{fig2cloudoct31}, but showing the fitted cloud parameters for storm region for second set of 11$^{th}$ November observations using the two-layer Model A.  \label{fig2cloudnov11}}
\end{figure}

\begin{figure}
\epsscale{.80}
\plotone{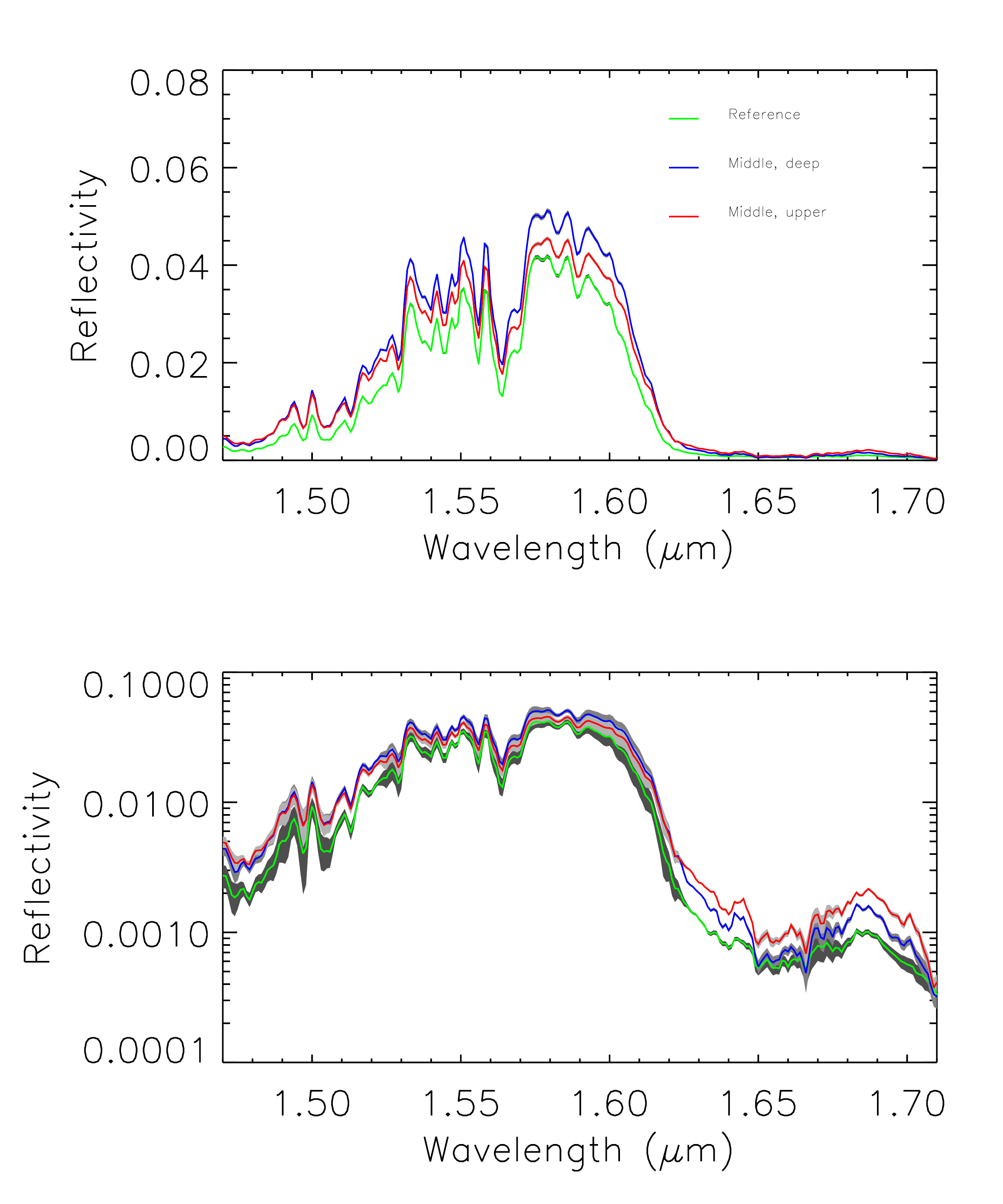}
\caption{As Fig.\ref{specompoct31}, but showing a comparison of the three sample observed spectra for second set of 11$^{th}$ November observations indicated in Fig.\ref{figregionselect}: (1) reference point away from storm; (2) middle of deeper cloud feature; (3) middle of upper cloud feature. The measurement errors estimated from the VLT/SINFONI pipeline are indicated by the grey regions. \label{specompnov11}}
\end{figure}

\begin{figure}
\epsscale{.80}
\plotone{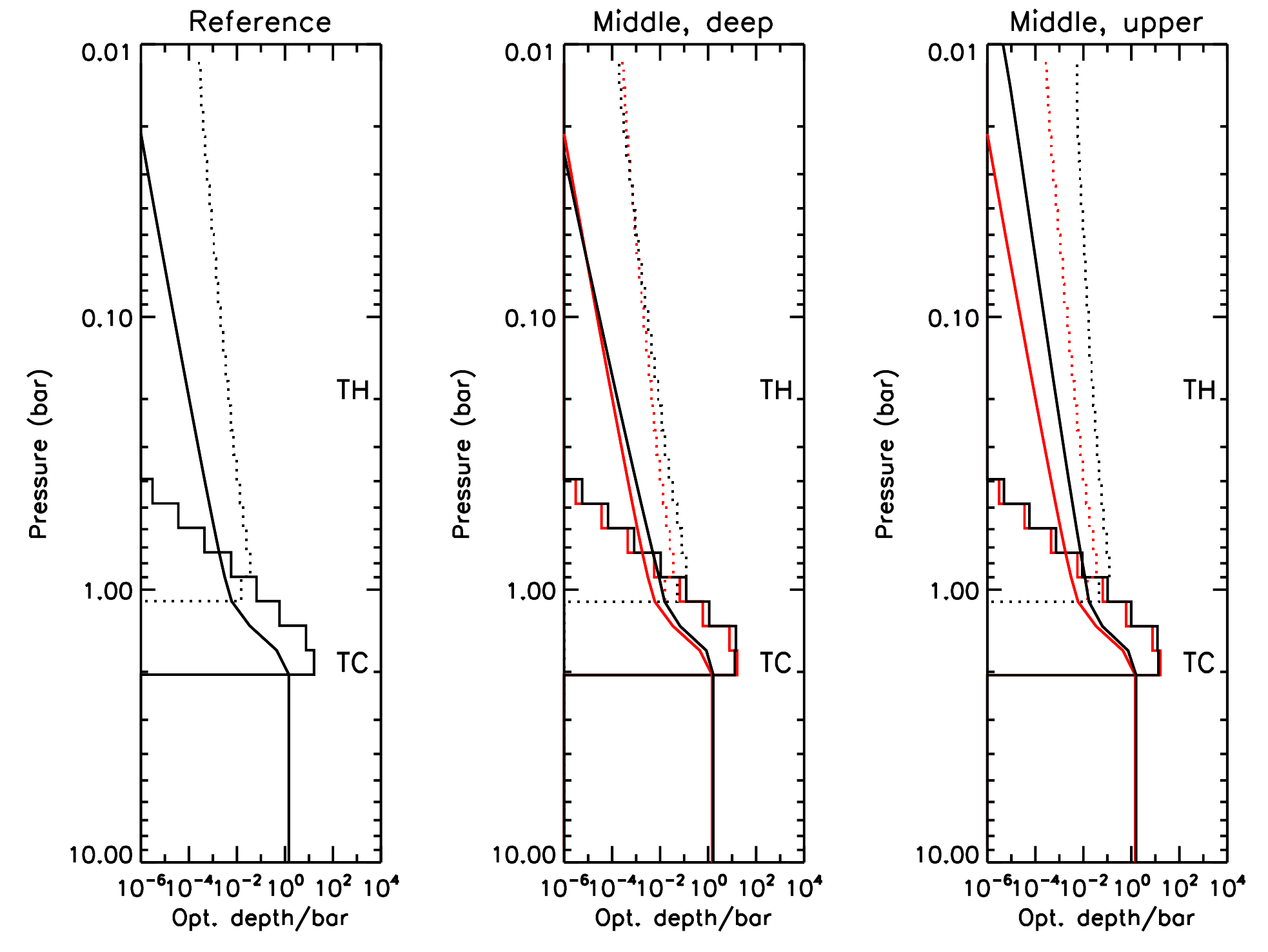}
\caption{As Fig.\ref{retrievedcloud2}, but showing a comparison of three fitted cloud profiles (using the two-layer Model A) for the second set  of 11$^{th}$ November observations: (1) reference point away from storm; (2) middle of deeper cloud feature; (3) middle of upper cloud feature. Optical depths are again quoted at a wavelength of 1.6 $\mu$m and the continuous line increasing with pressure in all panels is the integrated optical depth. Once again the retrieved profiles from the reference point (left hand panel) are reproduced in red in the middle and right hand panels to aid comparison.\label{retrievedcloud2nov11}}
\end{figure}

We then applied our retrieval models to the observations made on 11$^{th}$ November 2014. We chose the second set on this night as this had the storm closer to the centre of the disc and thus the geometry matched more closely the 31$^{st}$ October retrievals and there was again less likelihood of systematic error due to mixing with space/near-limb pixels. Here (Fig. \ref{figsumfalsecol}) we see the same shift in the centre of the storm  as we go from spectral regions of low to high methane absorption, but in this observation, taken nearly two weeks after those on 31$^{st}$ October, we can see that the upper component of the cloud is notably thicker at higher altitudes, but less thick at the base, making the `yellowish' component in Figs. \ref{figsumfalsecol} -- \ref{figfalsecolsingle} brighter and the `reddish' component dimmer. We applied both models to a smaller region about the storm cloud, indicated in Fig.\ref{figregionselect}, and the results for the two-layer Model A are presented in Fig. \ref{fig2cloudnov11}. A smaller region was chosen compared with the 31$^{st}$ October observations as the `tail' was not as clearly discernible and wide-area multiple scattering multivariate retrievals are, as mentioned earlier, computationally very expensive. Here we find that, once again, the deeper part of the cloud coincides with a brightening and very slight rising of the TC, while the upper part of the cloud coincides with increased opacity of the TH. The retrieved fractional scale height of the TH is again relatively featureless, but is once more lower in the vicinity of the storm cloud, where the opacity is greater. To see what this means in terms of cloud density as a function of height, Fig. \ref{specompnov11} compares the measured spectra at three points, indicated in Fig.\ref{figregionselect}: 1) a spot well away from the storm; 2) a point in the centre of the deep cloud; and 3) a point in the centre of the upper component of the cloud. In Fig. \ref{specompnov11} we can see the same trend in reflectivity spectra as seen on 31$^{st}$ October, although the spectrum from the deep cloud centre is noticeably dimmer, while both the deep- and intermediate-component cloud pixels are generally brighter at wavelengths of strong methane absorption. Figure \ref{retrievedcloud2nov11} compares the retrieved cloud profiles from these three points with the two-layer Model A. Here we can see that the TH is thicker and noticeably more vertically extended than for the equivalent retrievals on 31$^{st}$ October retrievals, which is consistent with our expectations from viewing the false colour images (Fig. \ref{figsumfalsecol}). 

\begin{figure}
\epsscale{.80}
\plotone{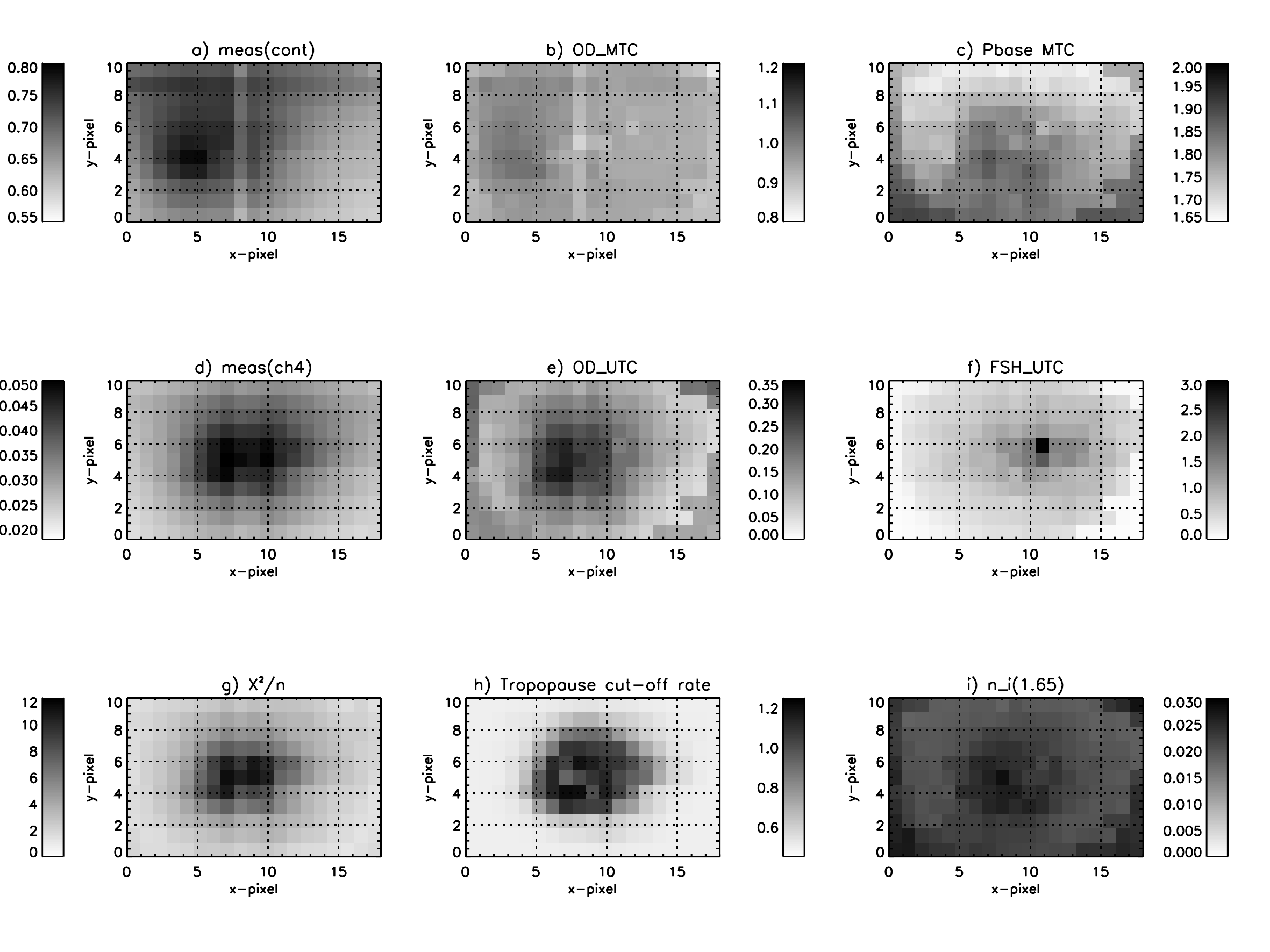}
\caption{As Fig.\ref{fig5cloudoct31}, but showing the fitted cloud parameters for storm region for second set of 11$^{th}$ November observations using the 5-layer Model B. Note here that the opacity of the TH was fixed to 0.002 to prevent the retrieval becoming unstable.\label{fig5cloudnov11}}
\end{figure}

\begin{figure}
\epsscale{.80}
\plotone{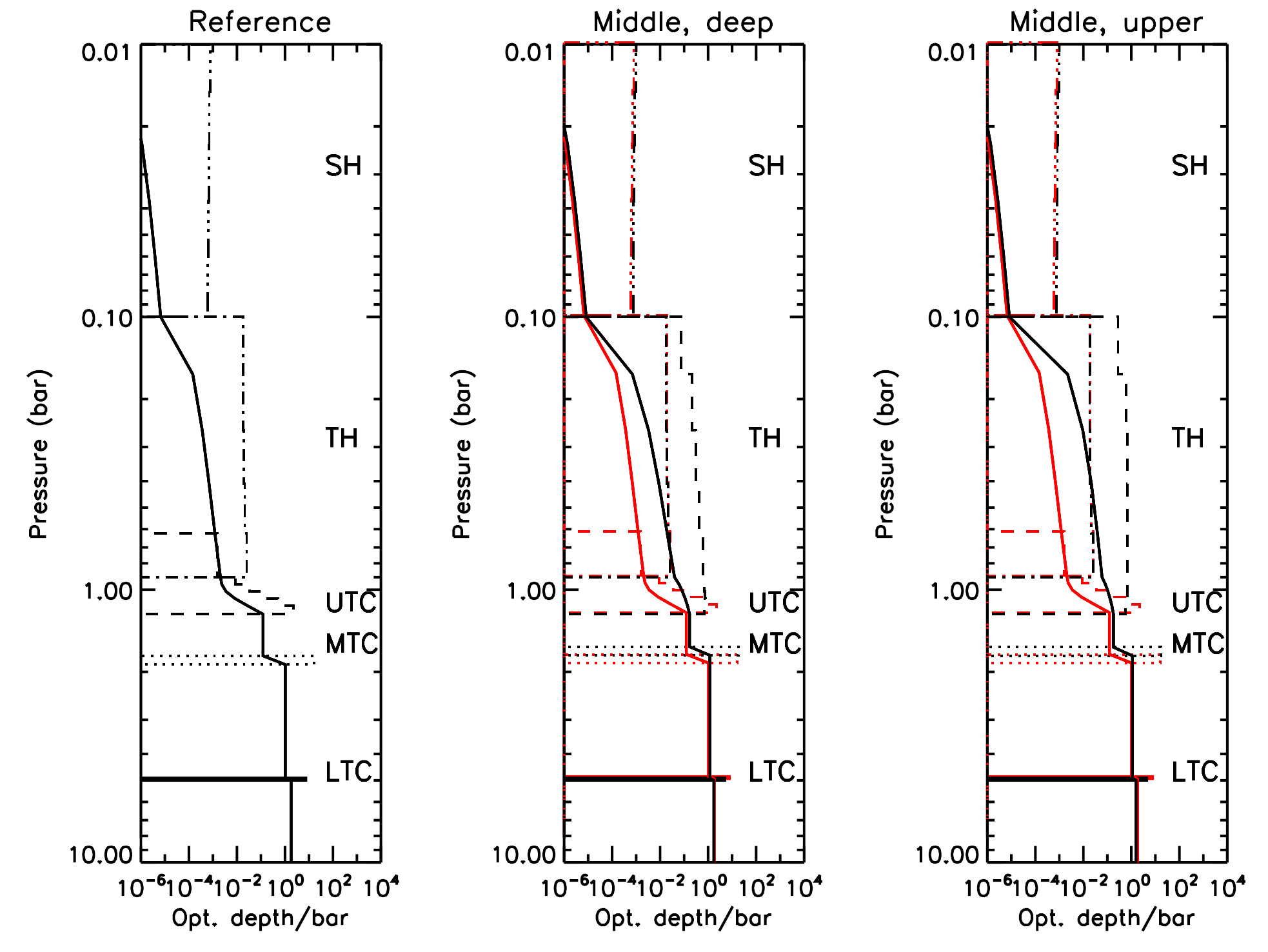}
\caption{As Fig.\ref{retrievedcloud5}, but showing a comparison of three fitted cloud profiles (using the five-layer Model B) for the second set  of 11$^{th}$ November observations: (1) reference point away from storm; (2) middle of deeper cloud feature; (3) middle of upper cloud feature. The similarity in the vertical distribution of the UTC and TH for Point 3 leads to the TH opacity becoming indistinguishable from the UTC parameters  at pixels near the centre of the upper level cloud, causing the instability that forced us to fix the TH opacity in the $11^{th}$ November retrievals. Once again the retrieved profiles from the reference point (left hand panel) are reproduced in red in the middle and right hand panels to aid comparison.\label{retrievedcloud5nov11}}
\end{figure}

The results for the more complicated five-layer Model B for the 11$^{th}$ November observation are presented in Fig. \ref{fig5cloudnov11}. Here we find that the deeper part of the cloud again coincides with thickening of the MTC (although the correlation is much less clear than for the 31$^{st}$ October observations), while the upper part coincides with increased opacity of the UTC, rather than an increase in the UTC's fractional scale height, which is generally much higher everywhere than was seen on 31$^{st}$ October. Instead, the upper part of the cloud coincides better with the high tropopause cut-off rates needed to limit the UTC's opacity near the tropopause. Please note that the scales on Fig.\ref{fig5cloudnov11} are different from those used in Fig. \ref{fig5cloudoct31}. Figure \ref{retrievedcloud5nov11} compares the retrieved cloud profiles from these three points. Comparing Fig.\ref{retrievedcloud5nov11} with Fig.\ref{retrievedcloud5} we can see that the UTC is generally thicker and that the retrieval is trying to shift the main position up from the fixed base at 1.23 bars. Our NEMESIS retrieval code achieves this by retrieving a fractional scale height in the cloud greater than 1.0, which means that the cloud density falls less quickly with height than pressure and thus the relative abundance of cloud increases. However, such an increase cannot continue indefinitely or we would end up with significant amounts of the UTC in the stratosphere, which is not thought allowed by this model, and which would also lead to significant reflectance at the wavelengths of strongest methane absorption, which is not seen \citep{deKleer15}. Hence, NEMESIS increases the tropopause cut-off parameter to bring the number density more steadily back to zero at the tropopause (0.1 bar). This considerably extended UTC makes the signature of the UTC and TH virtually inseparable and initial retrievals became unstable. Hence, in the retrievals presented here the opacity of the TH was fixed to the mean value determined from the 31$^{st}$ October observations of 0.002. 

\begin{figure}
\epsscale{.80}
\plotone{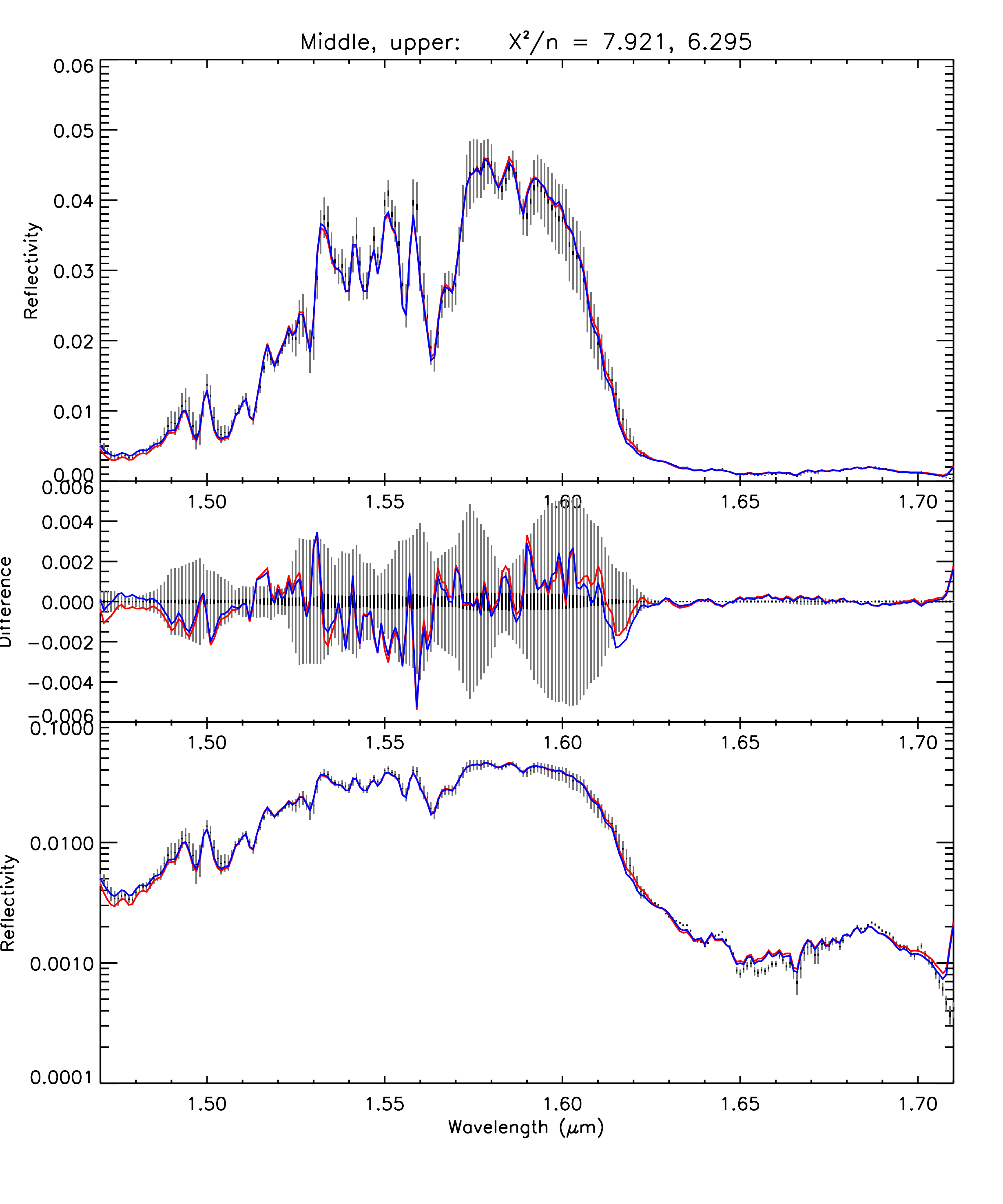}
\caption{As Fig.\ref{specref3oct31}, but comparing the observed and fitted spectrum at point 3 in the middle of the upper cloud feature for 11$^{th}$ November (indicated in Fig.\ref{figregionselect}), where our fit is poorest. Again, the blue line is the spectrum computed with the 2-layer model, while the red line is that calculated with the 5-layer model. Again, for the error bars, the lighter grey lines indicate measurement plus forward modelling errors, while the black error bars indicate the measurement errors alone as estimated by the VLT/SINFONI pipeline. The $\chi^2/n$ values in the title again indicate the closeness of fit for the 5-layer and 2-layer models respectively.\label{specref3nov11}}
\end{figure}

As before we find that both Models A and B match the data reasonably well for the 11$^{th}$ November observation, with the simpler Model A doing slightly better. Although the final $\chi^2/n$ values look large in some regions, especially at pixels in the centre of the upper cloud component, we actually achieve a very good fit with both models as can be seen in Fig.\ref{specref3nov11}. The large values of $\chi^2/n$  seen actually have more to do with the very low estimated forward model error of \cite{tice13} at 1.63 -- 1.65 $\mu$m than inherent model inadequacies, although both models have difficulty in fitting the 1.65 to 1.67 $\mu$m region, suggesting there is too much cloud at upper levels in the model. For Model A this is likely because the TH is free to extend past the tropopause while for Model B this may partly have been caused by our fixing the TH opacity in this retrieval to ensure stability, but more likely indicates that our chosen three-parameter model of the vertical distribution of UTC opacity is too restrictive in that the base pressure is fixed.

\section{Discussion}

The centre of the cloud observed is at a latitude (planetocentric) of 34.5$^\circ$. At this latitude, the zonal wind strength on Uranus is estimated to be $\approx 100$m/s in the prograde direction \citep{hammel01} and this is a region of maximum horizontal wind-shear (i.e. maximum $du/dy$). This may help to confine the feature into a `vortex' and explain its longevity and that of other cloud features that regularly appear at this latitude. From the temperature fields determined by Voyager 2 \citep{conrath98} Uranus' winds are estimated to be decaying with height and more recent ground based observations of Uranus' thermal structure suggest there has been no ostensible change since the Voyager 2 observations \citep{orton15}. From its morphology, the cloud appears to be a convective event, formed deep in the atmosphere, which then `bubbles up'  into the high atmosphere where the winds are slightly lower, causing the cloud column to become sheared with height in the direction we see. We were initially concerned that this vertical wind-shear would soon disrupt this cloud and spread it along this latitude band, rather as was seen for Saturn's northern springtime storm disturbance, or `Great White Spot', observed in 2010/2011 \citep{sanchezlavega11, fletcher11, fletcher12}. However, this latitude band (30 -- 45$^\circ$N) is also one where there is a minimum in upper tropospheric temperatures and thus $\partial T/\partial y$ approaches zero. From the Thermal Wind Equation we expect $\partial u / \partial p \propto \partial T/\partial y$ and thus the vertical wind shear should be at a minimum at this latitude, which would also help to explain the cloud's longevity. Assuming the feature is indeed a convective event, confined by the zonal winds as described, the `tail' of material immediately to the south, which is seen in these observations and also the Keck observations of `Feature 2' of \cite{dePater15}, would appear to be cloud material that is somehow escaping from this vortex and trailing along in the slower moving zonal winds immediately to the feature's south. 

An alternative interpretation of this feature is that we are instead seeing some sort of wave feature, where air rises and falls through the disturbance as it flows zonally (perhaps akin to the process of clouds formed by orographic uplift), and the wave propagation characteristics lead to the disturbance's vertical structure. Although for the 5-layer Model B we keep the UTC fixed at 1.23 bar, we allowed the base pressure of the MTC to vary and we can see that this moves to lower pressures in the deep cloud centre, which is consistent with this picture. It is also possible that allowing the base of the UTC to move would improve the fit to the observed spectra, but if this cloud is really a methane condensation cloud that might imply either a reduction in the methane mole fraction from 4\% to something lower, or instead a local increase in temperature. Intriguingly, such an increase in temperature might be consistent with the latent heat released by the apparent thick cloud condensation. However, there are other ways in which the base pressure of the UTC might be variable: a lack of condensation nuclei, for example, might cause such an effect and, in addition, we did not consider here the effects of methane depletion (which was explored by \cite{deKleer15}), which could contribute to a variable methane cloud base pressure. From its size (roughly 3,500 km in diameter) the feature is too large to be explained simply by gravity waves. We considered the possibility that it is the manifestation of a Rossby wave; however, if this were the case we would expect to see a semi-regular train of features along this latitude. While it is conceivable that the faint features seen $\sim 60^{\circ}$ away in the retrograde direction might be a manifestation of such a wave, on the whole, we find the properties of the feature more consistent with a localized convective event, which evolved during our observation sequence, becoming more vertically extended. Observations at higher SNR would allow the fainter features to be probed in order to better investigate their relationship with the brighter feature studied here.

Assuming it is a convective cloud feature it is interesting to consider how such an event might be triggered. At the high temperatures in Uranus' deep atmosphere we expect the ratio between the ortho-H$_2$ and para-H$_2$ isomers to be 3:1, equivalent to a para-H$_2$ fraction ($f_p$) of 25\%. At lower temperatures, and given sufficient time to come to equilibrium the para-H$_2$ fraction ($f_p$) rises. Hence, regions of low $f_p$ can be considered to be regions where air has risen from below faster than the para-H$_2$ isomer can be converted into ortho-H$_2$. Together with being the coldest upper tropospheric region, the 30 -- 45$^\circ$N latitude range is also one with the lowest para-H$_2$ fraction compared with the rest of the northern hemisphere \citep{orton15}, which suggests that this region is one where material is rising up from below. This view is consistent with the earliest interpretations of Voyager IRIS observations \citep{flasar87}, which suggested that the cool temperatures seen in the upper troposphere at these latitudes were caused by upwelling, with corresponding downwelling at the equator and poles. \cite{orton15} show that the distribution of upper tropospheric temperatures detected by Voyager 2 has not changed noticeably in the intervening years. However, this view is less consistent with microwave observations of the latitudinal deep abundance of ammonia (e.g. \cite{orton07}) and near-IR observations of the latitudinal abundance of methane (from $\sim 1$ -- 3 bars) \citep{kark09,tice13,sromovsky14}, which show the air to be `moist' from $45^\circ$S to $45^\circ$N and dry elsewhere, suggesting upwelling at equatorial latitudes and subsidence towards the poles (i.e. the 1-cell model of \cite{sromovsky14}). Such a flow is also not consistent with the vertically stacked 3-cell circulation model of \cite{sromovsky14}, modified to better account for poleward and equatorward flow of clouds at different altitudes, but it is worth noting that neither of these models appears well constrained at pressures less than 1 bar. It may be that the general equatorial-upwelling / polar-downwelling flow weakens near the equator in the upper troposphere, giving rise to the circulation suggested by mid-infrared observations evidenced by low temperatures and low $f_p$. However, unlike other examples of low-$f_p$ regions on the Giant Planets, such as the Equatorial Zones on both Jupiter and Saturn, where a thick cloudy zone is formed, it appears that on Uranus, this upwelling is somehow temporarily `trapped' below the main cloud deck at 2 -- 3 bar, before being occasionally released into large convective features such as the cloud feature observed here.  In other words, it is possible that there might be a build up of Convective Available Potential Energy (CAPE) beneath the cloud deck, followed by occasional `eruptions', rather like Saturn's storm or Jupiter's SEB revival.  Such events could be seasonally triggered and it is intriguing that these large storms are being seen in the springtime hemispheres of both Uranus and Saturn.  It is possible that the increased seasonal warming changes the stability of the overlying troposphere with regards to moist convective penetration on both planets, potentially triggering these events.

\section{Conclusions}

Our VLT/SINFONI observations of the bright storm cloud detected in Uranus' atmosphere in the autumn of 2014 contain unique information on the vertical structure and temporal evolution of this cloud feature, allowing much finer vertical resolution than can be achieved by filter-imaging observations alone, albeit at slightly poorer spatial resolution. We analysed these observations with our radiative transfer and retrieval code, NEMESIS using a 2-layer model (Model A) preferred by \cite{tice13} and also a version of the five-layer Uranus cloud scheme (Model B) proposed by \cite{sromovsky11}, modified to include a vertically extended Upper Tropospheric Cloud (UTC), which is assumed to be composed of methane ice. Both models fit the observations equally well, but assuming that this cloud is a methane ice condensation feature, we conclude from Model B that during the period of observations the storm feature was well modelled by variations of two of the main component clouds: 1) a brightening (i.e. an increase in reflectivity) and rising of the main Middle Tropospheric Cloud (MTC) deck at the 2 -- 3 bar level; and 2) a thickening and vertical extension of the Upper Tropospheric Cloud (UTC), based at 1.23 bar and assumed to be composed of methane ice. Our observations show that the centre of storm cloud shifts markedly with increasing altitude, with the top of the cloud centred significantly further upwind than the base. From its morphology the feature appears to be an ephemeral condensed cloud, presumably formed by convective upwelling, and we find that the latitude band in which the feature sits (30 -- 40$^{\circ}$N) may be particularly conducive to the longevity of such features as it is a region of maximum latitudinal, but minimum vertical wind shear.  

Between 31$^{st}$ October and 11$^{th}$ November 2014 and assuming the 5-layer Model B we found that the UTC (i.e. methane cloud) component thickened significantly and became more vertically extended, while the MTC (i.e. main cloud at 2 --3 bar) component faded. The `tail' feature seen to the south of the main cloud in these observations (but which was less apparent on 11$^{th}$ November), and also by \cite{dePater15}, was accounted for by a thickening of the UTC, which was also responsible for the brighter polar `hood' seen in all observations polewards of $\sim 45^\circ$N.

Our simple three-component model (optical depth, fractional scale height, cut-off parameter) of the UTC was found to give good results everywhere, except in the very middle of the upper cloud, especially in the second set of observations on 11$^{th}$ November 2014. It would appear that in these cases the model requires refinement, perhaps by allowing the UTC base pressure to vary, which will be the subject of further work. This further work will also incorporate the higher zenith-angle observations taken of the same cloud on both nights, which could improve upon the vertical resolution. We find that to place meaningful constraints on the opacity variations of the tropical and stratospheric hazes requires simultaneous measurements over a wider spectral range of the same feature in order to disentangle the contributions of the different layers in the Uranus vertical cloud scheme of \cite{sromovsky11}. Such measurements will have to wait until the next bright cloud of this type is spotted in Uranus' atmosphere and can be observed simultaneously over a wider spectral range.

Finally, these observations show the efficacy of modern Integral Field Unit spectrometers such as VLT/SINFONI in probing the vertical cloud structure of Uranus and the other giant planets. Although Uranus is long past its equinox in 2007, its atmosphere remains highly active. It will be of great interest to see how this activity evolves and whether there will be any further convective outbreaks as Uranus' north pole swings towards the Sun over the next decade until it reaches northern winter solstice in 2028. By that stage, it will presumably have reverted to the subdued, quiescent state observed by Voyager 2 during its encounter with Uranus at southern summer solstice in 1986.






\section{Acknowledgements}

We are very grateful to our VLT/SINFONI support astronomer George Hau who helped us hugely in designing these observations at short notice and executing them. Leigh Fletcher was supported by a Royal Society Research Fellowship at the University of Oxford. Glenn Orton was supported by a grant from NASA to the Jet Propulsion Laboratory, California Institute of Technology. The VLT/SINFONI observations were performed at the European Southern Observatory (ESO), Proposal 294.C-5004, as part of the Director's Discretionary Time (DDT). We thank an anonymous referee for very detailed and thoughtful comments, which helped to considerably improve our paper. 



{\it Facilities:} \facility{VLT (SINFONI)}.

\clearpage



\clearpage

\clearpage

\clearpage

\clearpage

\clearpage

\clearpage

\begin{deluxetable}{llccccccccccccc}
\tablewidth{0pt}
\tablecaption{2014 VLT/SINFONI H-Grism Observations.\label{tbl-1}}
\tablehead{
\colhead{Date\tablenotemark{a}}             & \colhead{Target} & \colhead{$T_{start}$\tablenotemark{b}}      &
\colhead{$T_{end}$\tablenotemark{b}}            &
\colhead{N$_{exp}$}     & \colhead{T$_{exp}$}          & \colhead{NDIT\tablenotemark{c}}    &
\colhead{Scale} & \colhead{Airmass} & \colhead{Seeing}}
\startdata
20141031   & HD212874 & 01:27 & 01:27 & 1 & 10s & 2 & 0.1\arcsec & 1.176 & 1.2\arcsec \\
20141031  & Uranus\tablenotemark{e} & 01:42 & 02:17  & 24\tablenotemark{d} & 60s & 1 & 0.1\arcsec & 1.211 -- 1.164 & 1.05\arcsec \\
20141031\tablenotemark{f} & Uranus  & 02:21 & 02:56  & 24 & 60s & 1 & 0.1\arcsec & 1.161 -- 1.147 &  0.92\arcsec \\
\tableline
20141111 & Uranus & 00:36 & 01:12 & 24 & 60s & 1 & 0.1\arcsec & 1.257 -- 1.185 & 2.16\arcsec \\
20141111 & Uranus & 01:15 & 01:50 & 24 & 60s & 1 & 0.1\arcsec  & 1.183 -- 1.15 & 1.44\arcsec \\
20141111 & HD210780 & 02:00 & 02:00 & 1 & 4s & 2 & 0.1\arcsec & 1.2 & 1.16\arcsec \\

\enddata
\tablenotetext{a}{Dates are listed as YYYYMMDD}
\tablenotetext{b}{Times are UT}
\tablenotetext{c}{Number of Detector Integration Times (NDIT)}
\tablenotetext{d}{The observation sequence for Uranus combined four sets of observations, in which for each there were five planet observations (2$\times$ 2 mosaic plus once in the centre) and one sky observation.}
\tablenotetext{e}{Uranus sub-Earth, sub-solar latitudes were 26.65$^\circ$ and 27.82$^\circ$ respectively on October 31$^{st}$ and 26.28$^\circ$ and 27.94$^\circ$ on November 11$^{th}$. The angular diameters of Uranus on these dates were 3.69\arcsec and 3.67\arcsec respectively. }
\tablenotetext{f}{Observations were also made November 8$^{th}$ and 9$^{th}$ 2014 during poorer weather conditions, but these are of too low quality to present here.}
\end{deluxetable}

\clearpage

\begin{deluxetable}{llll}
\tablewidth{0pt}

\tablecaption{Five-layer Model B cloud scheme based on \cite{sromovsky11}  and associated cloud scattering properties.\label{tbl-2}}
\tablehead{
\colhead{Cloud}             & \colhead{P$_{base}$ (bar)} & \colhead{P$_{top}$ (bar)}      &
\colhead{Scattering Properties}}
\startdata
Stratospheric Haze & 0.1 & 0.01 & $n_{r}$ = 1.4, $n_{i}(\lambda) = 0.0055exp[(350 - \lambda)/100]$   \\
 (SH) & & & ($\lambda$ in nm) extrapolated from 1 to 2 $\mu$m.  \\
 & & & Mie scattering, assuming standard Gamma \\
 & & & distribution of sizes with $r_{0} = 0.1 \mu$m and\\
 & & &  variance = 0.3.\\
 \tableline
 Tropospheric Haze & 0.9 & 0.1 & $n = 1.4+0.001i$.  Mie scattering with \\
 (TH) & & & Gamma distribution  $r_{0} = 0.1 \mu$m, variance = 0.05 \\
 \tableline
 Upper Tropospheric & 1.23 & Extended\tablenotemark{a} & Refractive indices of \cite{martonchik94}.   \\
 Cloud (UTC) & & & Mie scattering with Gamma distribution   \\
 & & & $r_{0} = 1.2 \mu$m, variance = 0.1 \\
\tableline
Middle Tropospheric & P$_{base}$\tablenotemark{b}  & P$_{base}\times 0.93$ &  \textit{a priori} $n = 1.4+0.001i$.  Mie scattering with \\
Cloud (MTC) & (variable) & & Gamma distribution  $r_{0} = 1.0 \mu$m, variance = 0.05 \\
\tableline
Lower Tropospheric & 5 & 4.9 & Empirical Properties of TC derived by \\
Cloud (LTC) &  & $(5\times 0.98)$ & \cite{tice13} \\

\enddata
\tablenotetext{a}{The UTC extends up to the tropopause with a vertical profile defined by a fractional scale height and cut-off parameter as described in the text.}
\tablenotetext{b}{P$_{base}$ of MTC was fitted by retrieval model.}
\end{deluxetable}

\clearpage

\end{document}